\documentclass[pra,aps,twocolumn,showpacs,superscriptaddress,nofootinbib]{revtex4}

\input{Qcircuit}
\input epsf
\renewcommand{\Qcircuit}[1][0em]{\xymatrix @*[o] @*=<#1>}
\newcommand{\node}[2][]{{\begin{array}{c} \ _{#1}\  \\ {#2} \\ \
\end{array}}\drop\frm{o} }
\newcommand{\link}[2]{\ar @{-} [#1,#2]}

\begin{document}

\title{Noise thresholds for optical cluster-state quantum computation}

\author{Christopher M. Dawson}
\affiliation{School of Physical Sciences,
The University of Queensland, Queensland 4072, Australia}

\author{Henry L. Haselgrove}
\email{HLH@physics.uq.edu.au} \affiliation{School of Physical
Sciences, The University of Queensland, Queensland 4072, Australia}
 \affiliation{Information
Sciences Laboratory, Defence Science and Technology Organisation,
Edinburgh 5111 Australia}

\author{Michael A. Nielsen}
\email[http://www.qinfo.org/people/nielsen/blog/]{}
\affiliation{School of Physical Sciences,
The University of Queensland, Queensland 4072, Australia}

\date{\today}

\begin{abstract}
  In this paper we do a detailed numerical investigation of the
  fault-tolerant threshold for optical cluster-state quantum
  computation.  Our noise model allows both photon loss and depolarizing
  noise, as a general proxy for all types of local noise other than
  photon loss noise.  We obtain a \emph{threshold region} of allowed
  pairs of values for the two types of noise.  Roughly speaking, our
  results show that scalable optical quantum computing is possible in the combined presence of both
  noise types, provided that the loss probability is $< 3 \times 10^{-3}$
  and the
  depolarization probability is $< 10^{-4}$.  Our fault-tolerant
  protocol involves a number of innovations, including a method for
  syndrome extraction known as telecorrection, whereby repeated
  syndrome measurements are guaranteed to agree.  This paper is an
  extended version of [Dawson et al., Phys.~Rev.~Lett.~{\bf 96}, 020501].
\end{abstract}

\pacs{03.67.-a,03.67.Lx}

\maketitle

\section{Introduction}

Optical systems have many significant advantages for quantum
computation, such as the ease of performing single-qubit
manipulations, long decoherence times, and efficient read-out.
Unfortunately, standard linear optical elements alone are unsuitable
for quantum computation, as they do not enable photons to interact.
This difficulty can, in principle, be resolved by making use of
nonlinear optical elements~\cite{Yamamoto88a,Milburn89a}, at the price
of requiring large nonlinearities that are currently difficult to
achieve.

An alternate approach was developed by Knill, Laflamme and Milburn
(KLM)~\cite{Knill01a}, who proposed using measurement to effect
entangling interactions between optical qubits.  Using this idea, KLM
developed a scheme for scalable quantum computation based on linear
optical elements, together with high-efficiency photodetection,
feed-forward of measurement results, and single-photon generation.
KLM thus showed that scalable optical quantum computation is in
principle possible. Experimental
demonstrations~\cite{Pittman03a,OBrien03a,Sanaka03a,Gasparoni04a,Zhao05a}
of several of the basic elements of KLM have been achieved.

Despite these successes, the obstacles to fully scalable optical
quantum computation with KLM remain formidable.  The biggest challenge
is to perform a two-qubit entangling gate in the near-deterministic
fashion required for scalable quantum computation.  KLM propose an
ingenious scheme showing that this is possible in principle, but with
a considerable overhead: doing a single entangling gate with high
probability of success requires tens of thousands of optical elements.
Several proposals (e.g.,~\cite{Franson02a,Gilchrist05a}) have been
made to reduce this overhead, but it still remains formidable even in
these improved schemes.

A recent proposal~\cite{Nielsen04b} (c.f.~\cite{Yoran03a}) combines
the basic elements of KLM with the cluster-state model of quantum
computation~\cite{Raussendorf01a} to achieve a reduction in complexity
of many orders of magnitude.  This scheme has been further simplified
in~\cite{Browne05a}, where it is estimated that only tens of optical
elements will be required to implement a single logical gate.  The
resulting proposal for optical cluster-state quantum computing thus
appears to offer an extremely promising approach to quantum
computation.  Recent experiments~\cite{Walther05a} have demonstrated
the construction of simple optical cluster states.  A recent review of
work on optical quantum computation, including cluster-based
approaches is~\cite{Kok05a}.

While the optical cluster-state proposals~\cite{Nielsen04b,Browne05a}
present encouraging progress, for them to be considered credible
approaches to fully scalable quantum computation, it is necessary to
consider the effects of noise. In particular, it is necessary to
establish a \emph{noise threshold theorem} for the optical
cluster-state proposals. A noise threshold theorem proves the
existence of a constant \emph{noise threshold} value, such that
provided the amount of noise per elementary operation is below this
level, it is possible to efficiently simulate a quantum computation of
arbitrary length, to arbitrary accuracy, using appropriate
error-correction techniques.  In the standard quantum circuit model of
computation such a threshold has been known to exist since the
mid-1990s (see Chapter~10 of \cite{Nielsen00a} for an introduction and
references).  However, the optical cluster-state proposals are not
based on the circuit model, but rather on the cluster-state model of
computation, and thus \emph{a priori} it is not obvious that a similar
noise threshold need hold.

Fortunately, recent work~\cite{Nielsen05a,Raussendorf03b,Aliferis05a}
has shown that the fault-tolerance techniques developed for the
circuit model can be adapted for use in the cluster-state model, and
used to prove the existence of a \emph{noise threshold} for noisy
cluster-state computing.  The earliest
work~\cite{Nielsen05a,Raussendorf03b} established the \emph{existence}
of a threshold for clusters, without obtaining a value.
\cite{Aliferis05a} argued that in a specific noise model, the cluster
threshold is no more than an order of magnitude lower than the
threshold for circuits.  The most recent work~\cite{Raussendorf05a}
combines ideas from cluster-state computing with topological
error-correction to obtain a cluster threshold.  However,
neither~\cite{Aliferis05a} nor~\cite{Raussendorf05a} are of direct
relevance to the optical cluster-state proposal, since they make use
of deterministic entangling gates, which are not available in linear
optics, and the noise model does not include any process analogous to
photon loss.

The present paper studies in detail the value of the noise threshold
for optical cluster state computing. We use numerical simulations to
estimate the threshold for a particular fault-tolerant protocol, for
two different quantum codes. The paper is an extended version of an
earlier work~\cite{Dawson06a}, which provided an overview (but few
details) of the protocol and simulation techniques used, and only a
brief summary of results.

Our threshold analysis is tailored to the dominant sources of noise in
optical implementations of quantum computing.  In particular, our
simulations involve three different sources of noise: (a) the inherent
nondeterminism of the entangling gates used to build up the cluster;
(b) photon loss; and (c) depolarizing noise.  The strength of noise
source~(a) is regarded as essentially fixed, while the strengths
of~(b) and~(c) are regarded as variables that can be changed by
improved engineering.  Note that most existing work on thresholds
(e.g.,~\cite{Aliferis05a,Raussendorf05a,Knill05a,Steane03a}) in either
clusters or circuits focuses on abstract noise models based on
depolarizing noise, and neglects sources~(a) and~(b).

Noise sources~(a) and~(b) likely dominate actual experiments, and our
protocol attempts to cope with these very efficiently. The protocols
for decoding and correction can be made to take advantage of the
knowledge the experimenter has of the locations of error types~(a)
and~(b). For example, the well-known Steane 7-qubit code is usually
used to correct a depolarization error on a single qubit. A more
efficient use of the code is possible, in which it is used to correct
photon loss or nondeterministic gate failure errors on as many as {\em
  two} qubits.

Although noise sources~(a) and~(b) will dominate, sources of noise
other than~(a) and~(b) will also be present in experiments, and so it
is important that our fault-tolerant protocol and analysis also deals
with those.  This is why we include noise source~(c), as a proxy for
all additional noise effects. Of course, in practice it is unlikely
that depolarizing noise will be a particularly good model for the
other noise sources.  However, standard results in the theory of
fault-tolerance imply that the ability to correct depolarizing noise
implies the ability to correct essentially all reasonable physical
noise models, and so depolarization is a good proxy for those other
effects.

A prior work~\cite{Varnava05a} (c.f.~\cite{Ralph05a}) has calculated a
threshold for optical quantum computation when the only source of
noise is photon loss.  In real experiments noise sources other than
photon loss are present, such as dephasing, and protocols such
as~\cite{Varnava05a,Ralph05a} will amplify the effects of such noise
at the encoded level.  Thus, even if the original noise strength is
very weak, encoding may amplify the noise to the point where it
dominates the computation.  By contrast, our protocol protects against
both photon loss and depolarizing noise, and by standard
fault-tolerance results thus automatically protects against arbitrary
local noise, including dephasing (in any basis), amplitude damping,
etc.

Because our model includes multiple noise parameters, we do not obtain
a single value for the threshold, as in most earlier work. Instead, we
obtain a threshold \emph{region} of noise parameters for which
scalable quantum computing is possible. The main outcome of our paper
is a series of threshold regions, with the different regions
corresponding to varying assumptions regarding the relative noise
strength of quantum memory, and the use of different quantum
error-correcting codes. Qualitatively, we find that our fault-tolerant
protocols are substantially more resistant to photon loss noise than
they are to depolarizing noise, with threshold values of approximately
$6 \times 10^{-3}$ for photon loss noise (in the limit of zero
depolarization noise), and $3\times10^{-4}$ for depolarizing noise (in
the limit of no photon loss).  When both types of noise are present in
the system, a typical value in the threshold region has a strength of
$3 \times 10^{-3}$ for photon loss noise, and a depolarization
probability of $10^{-4}$.

Our fault-tolerant protocol involves a number of innovations in
addition to those already described, including: (1) the development of
special techniques to deal with the inherent non-determinism of the
entangling optical gates; (2) heavy use of the ability to parallelize
cluster-state computations~\cite{Raussendorf03a}, and the ability to
do as much of the computation off-line as possible; and (3) as a
special case of the previous point, we develop a new method for doing
fault-tolerant syndrome measurement which we call
\emph{telecorrection}.  This has the striking property that repeated
measurements of the syndrome are \emph{guaranteed to agree} (which
helps increase the threshold), unlike in standard protocols, where
measurements only sometimes agree.

The structure of the paper is as follows.  In
Section~\ref{sec:background} we briefly overview the required
background on cluster-state computation and the optical cluster-state
proposal.  Section~\ref{sec:physical_setting} describes our
assumptions about the physical setting: what physical resources we are
allowed, what quantum gates we can perform, and what noise is present
in the system.  Section~\ref{sec:simulations} describes briefly how we
simulate noisy cluster-state computations.  This is a surprisingly
subtle topic, due to the multiple noise sources in our model, which is
why it merits a separate section.  Section~\ref{sec:protocol}
describes the details of the fault-tolerant protocol that we simulate,
and presents the results of our simulations, including threshold
regions for two different quantum codes.  Section~\ref{sec:conclusion}
concludes.

\section{Background}
\label{sec:background}

In this section we introduce the required background on cluster states
(Subsection~\ref{subsec:cluster_states}), and optical cluster-state
computation (Subsection~\ref{subsec:optical_cluster_states}).  The
main purpose is to fix notation and nomenclature, and the reader
looking for a more detailed introduction to these topics is referred
to, e.g.,~\cite{Nielsen05b,Raussendorf03a} for cluster states, and
to~\cite{Nielsen04b,Browne05a} for optical cluster-state computation.

\subsection{Cluster-state computation}
\label{subsec:cluster_states}

In this subsection we explain the cluster-state model of computation,
and how it can be used to simulate quantum circuits. (By ``simulate''
in this context, we are referring to a procedure for converting a
quantum circuit to an equivalent cluster-state computation). We give a
rather in-depth treatment of the simulation procedure here, as our
later discussion of fault-tolerance depends heavily on a thorough
understanding of the details of this procedure.  The presentation in
this subsection is based on the treatments
in~\cite{Nielsen05b,Nielsen05a}, which in turn are based
on~\cite{Raussendorf01a,Raussendorf03a}.  The reader is referred
to~\cite{Nielsen05a,Raussendorf03a} for proofs and a more in-depth
discussion.

We begin by explaining the cluster-state model itself, initially
ignoring the question of how a cluster-state computation can be used
to simulate a quantum circuit. Broadly speaking, a cluster-state
computation involves three steps: (1) the preparation of a special
entangled many-qubit quantum state known as a \emph{cluster state};
(2) an adaptive sequence of single-qubit measurements processing the
cluster qubits; and (3) read-out of the computation's result from the
remaining cluster qubits.  We now describe each step in detail.

The term ``cluster state'' refers not to a single quantum state, but
rather to a family of quantum states.  The idea is that an $n$-qubit
cluster state is specified by a graph on $n$ vertices; to each vertex
we associate a corresponding qubit in the cluster, and we apply a
graph-dependent preparation procedure to the qubits in order to define
the cluster (as described below).  For example, the following graph
represents a six-qubit cluster:
\begin{equation}
\Qcircuit[2em] @R=1em @C=1em {
\node{} & \node{} \link{0}{-1} \link{1}{0} & \node{} \link{0}{-1} \\
\node{} & \node{} \link{0}{-1}             & \node{} \link{0}{-1}
}
\end{equation}
The cluster state associated to such a graph may be defined as the
result of applying the following two-stage preparation procedure:
\begin{enumerate}
\item Prepare each of the $n$ qubits in the state $|+\rangle \equiv
  (|0\rangle+|1\rangle)/\sqrt 2$.

\item Apply {\sc cphase} (controlled-{\sc phase}) gates between
  cluster qubits whose corresponding graph vertices are connected by
  an edge.
\end{enumerate}
Although we have not specified the order in which the {\sc cphase}
gates in the second step are to be applied, this is okay, because
these gates all commute.  Note also that although this preparation
procedure defines the cluster state associated to the graph, it is of
course possible to use other preparation procedures to prepare the
same state. (An important case in point is discussed in the next
subsection -- how the so-called fusion gate may be used to prepare
optical cluster states).

Once the cluster is prepared, the next step in a cluster-state
computation is to perform a sequence of processing measurements on the
cluster.  These measurements are single-qubit measurements, whose
location and nature may depend (via a polynomial-size classical
computation) on the outcome of earlier measurements.

The output of the cluster-state computation, just before the final
readout, consists of (a) the quantum state $|\psi\rangle$ of the
qubits remaining after completion of the processing measurements, and
(b) the sequence of classical measurement outcomes obtained during
processing, which we denote $\mathbf{c}$.  These classical measurement
outcomes will generally affect the way in which we interpret the
quantum state output from the computation.  In particular, it is
convenient to regard the output as being the state
$\sigma_{\mathbf{c}} |\psi\rangle$, where $\sigma_{\mathbf{c}}$ is an
$n$-qubit product of Pauli matrices which is some suitable function of
the bit string $\mathbf{c}$.

It is often convenient to have a graphical representation of a
cluster-state computation.  For this purpose we use notation along the
following lines:
\begin{equation} \label{eq:first-cluster-figure}
\Qcircuit[4em] @R=1em @C=1em {
\node[1]{HZ_{\alpha_1}} & \node[2]{HZ_{\pm \alpha_2}} \link{0}{-1} \link{1}{0}
    & \node{} \link{0}{-1} \\
\node[1]{HZ_{\beta_1}}  & \node[2]{HZ_{\pm \beta_2}} \link{0}{-1}
    & \node{} \link{0}{-1}
}
\end{equation}
The overall shape of the diagram denotes the graph state to be created
at the beginning of the computation. Labels indicate qubits on which
processing measurements occur, while unlabeled qubits are those which
remain as the output of the computation when the processing
measurements are complete.  Note that qubits are labeled by a
positive integer $k$ and a single-qubit unitary, which we refer to
generically as $U$.  Here $U = HZ_{\pm
  \alpha_j}, HZ_{\pm \beta_j}$.  The $k$ label indicates the time
order in which processing measurements are to be performed.  Qubits
with the same label are allowed to be measured in any order relative
to each other, or simultaneously.  Time-ordering is important, because
some measurement results need to be fed-forward to control later
measurement bases. The $U$ label is used to specify the basis in which
the qubit is measured, indicating that the measurement may be
performed by applying the unitary $U$, and then performing a
computational basis measurement. (This is equivalent to performing a
measurement in the $\{ U^\dagger |0\rangle, U^\dagger |1\rangle\}$
basis.)  The $\pm$ notation in $HZ_{\pm \alpha_2}$ and $HZ_{\pm
  \beta_2}$ indicates that the choice of sign depends on the outcomes
of earlier measurements, in a manner that needs to be specified
separately.  Details of how this choice should be made are given later
in this subsection, and further examples can be found in,
e.g.,~\cite{Nielsen05a}.

We now describe a recipe that may be used to convert a quantum circuit
to a cluster-state computation. As part of this recipe we will
introduce the notion of a \emph{Pauli frame}. Initially the Pauli
frame will appear to be merely a bookkeeping device, but in our later
description of noisy cluster-state computation it will be an important
tool for tracking the effects of noise.

The key to simulating quantum circuits with cluster states is the
following circuit identity (\cite{Zhou00a}, see also
\cite{Nielsen05b})):
\begin{equation} \label{eq:transport}
\Qcircuit @C=1em @R=1.8em @!R {
 \lstick{|\psi\rangle} & \ctrl{1}     & \gate{HZ_\theta}
 & \meter & \cw & m \\
 \lstick{|+\rangle}    & \control \qw & \qw
 & \rstick{X^m H Z_\theta |\psi\rangle} \qw
}
\end{equation}
Note that the measurement basis is the computational basis, and $m =
0,1$ is the measurement outcome.  We shall call this circuit the
\emph{transport circuit}, since its effect is to transport (and
simultaneously transform) the quantum information input onto the
second qubit.

We will explain how to use cluster states to simulate quantum circuits
through a series of examples, starting with the following two-gate
single-qubit circuit:
\begin{equation} \label{eq:basic-one-qubit-circuit}
\Qcircuit @C=1em @R=1.8em @!R {
 \lstick{|+\rangle} & \gate{HZ_{\alpha_1}} & \gate{HZ_{\alpha_2}} & \qw
}
\end{equation}
Note that we assume the qubit starts in the $|+\rangle$ state, and
that single-qubit gates are of the form $HZ_\alpha$.  Later we will
show how to simulate multi-qubit circuits involving the {\sc cphase}.
Jointly, these operations are universal for computation, and so the
ability to simulate them is sufficient to simulate an arbitrary
quantum circuit.

The cluster-state computation used to simulate
Circuit~(\ref{eq:basic-one-qubit-circuit}) is:
\begin{equation} \label{eq:simple_simulating_cluster}
\Qcircuit[4em] @R=1em @C=1em {
 \node[1]{HZ_{\alpha_1}} & \node[2]{HZ_{\pm \alpha_2}} \link{0}{-1} &
\node{} \link{0}{-1}
}
\end{equation}
By definition, this cluster-state computation has an output equal to
the output of the following quantum circuit\footnote{Note that the
  double vertical lines emanating from the meter on the top qubit
  indicate classical feed-forward and control of later operations.}:
\begin{equation}
\Qcircuit @C=1em @R=1.8em @!R {
 \lstick{|+\rangle} & \ctrl{1}     & \gate{HZ_{\alpha_1}} & \meter &
\cw          & \cw \cwx[1] \\
 \lstick{|+\rangle} & \control \qw & \qw                  & \qw   &
\ctrl{1}     & \gate{HZ_{\pm \alpha_2}} & \meter \\
 \lstick{|+\rangle} & \qw          & \qw                  & \qw    &
\control \qw & \qw                      & \qw
\gategroup{1}{2}{2}{4}{1em}{--}
\gategroup{2}{5}{3}{7}{2.4em}{--}
}
\end{equation}
In this circuit we have measured the first qubit \emph{before} doing
the {\sc cphase} between the second and third qubits used during
creation of the cluster.  This does not change the output since these
operations are on different qubits, and thus commute.  We do this
because it enables us to understand the output as the result of two
cascaded circuits of the form of~(\ref{eq:transport}), as indicated by
the highlighted boxes.

It will be helpful to consider the quantum state of the qubits at each
of three intermediate locations.  The first location is the initial
state of the first qubit, i.e., $|+\rangle$.  Note that this is
exactly equal to the input for
Circuit~(\ref{eq:basic-one-qubit-circuit}).

The second location is the state of the second qubit output by the
first of the transport circuits and used as input to the second
transport circuit.  This state is $X^{m_1}H Z_{\alpha_1}|+\rangle$,
where $m_1$ is the output of the first measurement.  This is equal to
the state of the qubit in Circuit~(\ref{eq:basic-one-qubit-circuit})
after the first gate, up to the known Pauli matrix $X^{m_1}$.  We will
see shortly that we can compensate for this known Pauli matrix by a
suitable choice of the sign in $\pm \alpha_2$.

The third location is the state of the third qubit after both
transport circuits, i.e., at the end of the computation.  This state
is $X^{m_2} H Z_{\pm \alpha_2} X^{m_1} H Z_{\alpha_1}|+\rangle$, where
$m_2$ is the output of the measurement on the second qubit.  By
choosing the sign of $\pm \alpha_2$ so that $Z_{\pm \alpha_2} X^{m_1}
= X^{m_1} Z_{\alpha_2}$, and using the identity $H X^{m_1} = Z^{m_1}
H$, the output may be rewritten as $X^{m_2} Z^{m_1} H Z_{\alpha_2} H
Z_{\alpha_1}|+\rangle$.  Up to the known Pauli matrix $X^{m_2}
Z^{m_1}$, which can be compensated in post-processing, this is
identical to the output of the single-qubit
circuit~(\ref{eq:basic-one-qubit-circuit}).

The presence of these known Pauli matrices motivates us to define the
notion of a \emph{Pauli frame} for the cluster-state
computation~(\ref{eq:simple_simulating_cluster}), as follows.  At the
beginning of the computation, the Pauli frame is just the product $I
\otimes I \otimes I$ on three qubits.  We measure the first qubit,
with output $m_1$, and the updated Pauli frame is the two-qubit
operator $X^{m_1} \otimes I$.  We measure the second qubit, and the
updated Pauli frame is the single-qubit operator $X^{m_2}Z^{m_1}$.
Thus, at each stage the Pauli frame relates the state actually input
to the remaining stages of the cluster-state computation to the ideal
state of the circuit being simulated.

In general, suppose we are using a horizontal cluster to simulate an
arbitrary single-qubit computation.  Then by definition the initial
Pauli frame is just the tensor product of identities on all cluster
qubits.  Suppose at some stage we have a Pauli frame which is $X^x
Z^z$ on the first remaining cluster qubit, and acts as the identity on
all other qubits.  After measuring the first remaining cluster qubit,
and obtaining the result $m$, the updated Pauli frame is $X^{z+m} Z^x$
on the first remaining qubit after measurement, and the identity on
all the other qubits.

The Pauli frame is determined by the measurement results, and thus
will always be known to an experimenter observing the computation.
Furthermore, the Pauli frame determines the basis in which later
measurements are performed.  As the example above shows, if the Pauli
frame is $X^x Z^z$ on the qubit to be measured, then the measurement
basis to simulate a $HZ_\alpha$ gate is $HZ_{+\alpha}$ if $x = 0$, and
$HZ_{-\alpha}$ if $x = 1$.

The reader may wonder why we carry all the extra identity terms around
in the Pauli frame, since they aren't explicitly used.  Later we will
see that keeping these terms is quite useful in the analysis of noisy
cluster-state computations.

Let's generalize these ideas to the simulation of multi-qubit quantum
circuits.  Consider the following example, which illustrates the
general ideas:
\begin{equation}
\Qcircuit @C=1em @R=1.8em @!R {
 \lstick{|+\rangle} & \gate{HZ_{\alpha_1}} & \ctrl{1}     &
\gate{HZ_{\alpha_2}} & \qw \\
 \lstick{|+\rangle} & \gate{HZ_{\beta_1}}  & \control \qw &
\gate{HZ_{\beta_2}} & \qw
}
\end{equation}
This can be simulated using the cluster-state computation
of~(\ref{eq:first-cluster-figure}).  The correspondence between the
two is as follows.  Each qubit in the quantum circuit is replaced by a
horizontal row of cluster qubits.  As in the single-qubit case,
different horizontal qubits in the cluster represent the original
circuit qubit at different times, with each gate $HZ_{\alpha}$
replaced by a single qubit in the cluster.  {\sc cphase} gates in the
quantum circuit are simulated using a vertical ``bridge'' connecting
the appropriate cluster qubits in different rows.

As in the earlier example, at any given stage of the cluster-state
computation we define a Pauli frame which is a product of Pauli
operators on the remaining cluster qubits.  Initially, this is the
identity on all six cluster qubits.  Consider the measurement on the
two leftmost qubits.  In each case, the rule for updating the Pauli
frame is exactly as described earlier.  The only difference arises
when the qubit being measured has a vertical bond. In this case,
suppose prior to the measurement the Pauli frame has entry $X^{x_1}
Z^{z_1}$ on the qubit being measured, and entry $X^{x_2}Z^{z_2}$ on
the qubit attached via the vertical bond.  The update rule for after
the measurement is in two steps: (1) replace the Pauli frame on these
two qubits by $X^{x_1}Z^{z_1+x_2}$ and $X^{x_2}Z^{z_2+x_1}$,
respectively; (2) apply the earlier rules for horizontally attached
qubits, just as though the vertical bond was not present.

A generalization of the earlier analysis for the single qubit case
shows that with these rules, the state at the end of the cluster-state
computation is equal to the product of the Pauli frame with the output
from the quantum circuit being simulated.  Since the Pauli frame is
known by the experimenter, its presence can be compensated in
post-processing, and the two types of computation are equivalent.

We have described our simulations of circuits by a cluster-state
computation where the measurements are done left-to-right on the
qubits.  In fact, as emphasized in~\cite{Raussendorf03a}, when the
operations being simulated are Clifford group operations, no
measurement feed-forward is required, and it is possible to change the
order in which measurements are done.  In particular ``later'' parts
of the quantum circuit can be simulated earlier during the
cluster-state computation.  This means it is possible to
\emph{parallelize} operations that would be done at different times in
the circuit model, and even in some instances to \emph{premeasure}
parts of the cluster.  We will make heavy use of these ideas in our
fault-tolerant protocol, which consists entirely of Clifford group
operations.  Note however, that even in the cases where qubits may be
measured out of order, the update rules for the Pauli frame should be
applied as though measurements were done in the conventional
left-to-right order.

\subsection{Optical cluster-state computation}
\label{subsec:optical_cluster_states}

We've described cluster-state computation as an abstract model of
quantum computation.  We now discuss how cluster-state computation can
be implemented optically.  The method we use is a variant
of~\cite{Browne05a}, with key ideas coming from~\cite{Nielsen04b}
and~\cite{Knill01a}.

\begin{figure}[t]
\begin{center}
\setlength{\unitlength}{1cm}
\begin{picture}(3,4.5)
  \put(0,1.5){\vector(1,0){3}}  
  \put(1.5,0){\line(0,1){2.5}}  
  \put(1,1){\framebox(1,1)}     
  \put(1,1){\line(1,1){1}}      
  \put(1,2.5){\framebox(1,0.5){$45^\circ$}} 
  \put(1.5,3){\vector(0,1){0.5}} 
  \put(1,3.5){\line(1,0){1}}    
  \put(1.5,3.5){\oval(1,1.5)[t]}
\end{picture}
\end{center}
\caption{The fusion gate.  Two optical modes are combined on a polarizing
  beam splitter, which reflects vertically polarized light only.  The
  polarization of one mode is then rotated through $45^\circ$, before
  being measured using a polarization discriminating photon counter.
  \label{fig:fusion}}
\end{figure}
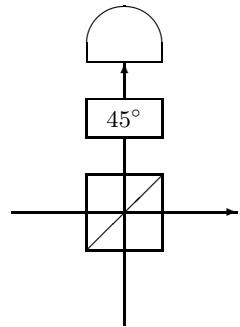

In this model, qubits are encoded in two optical polarizations
(horizontal $H$ and vertical $V$) corresponding to a single spatial
mode.  We will build up clusters using a supply of Bell states,
$(|HH\rangle+|VV\rangle)/ \sqrt 2$, and a gate known as the
\emph{fusion gate}, which is illustrated in Fig.~\ref{fig:fusion}.
Combining the ability to build up clusters with linear optics (used to
effect single-qubit rotations) and a polarization discriminating
photon counter, this thus enables quantum computing.  Note
that~\cite{Browne05a} uses two variants of the fusion gate; we will
only make use of one of these gates, and so refer to it simply as
\emph{the} fusion gate.

To see how this works, suppose a fusion gate is applied to two cluster
qubits that are not connected by an edge.  Then provided a single
photon is registered at the output (either horizontally or vertically
polarized), it can be shown the resulting quantum state is a cluster
state in which the two qubits have been \emph{fused}, i.e., combined
into a single cluster qubit whose edge set contains all the edges of
both fused qubits. This occurs with probability $50$ percent, and we
say that the fusion gate has been successful when it happens.  By
contrast, if zero or two photons register at the output, then the
fusion gate has failed, and it can be shown that the effect is to
delete the two qubits from the cluster. Failure occurs with
probability $50$ percent.

Using Bell states and fusion gates we can efficiently build up large
clusters.  The basic idea can be understood from the following example
of how to build up a linear cluster, following~\cite{Browne05a}.  Note
that the Bell state is simply a two-qubit cluster state, up to
unimportant local unitaries, and so we can prepare a pair of two-qubit
clusters:
\begin{equation}
\Qcircuit[2em] @R=1em @C=1em {
\node{} & \node{} \link{0}{-1} & \node{} & \node{} \link{0}{-1}}
\end{equation}
Successful fusion of a qubit from one Bell pair with a qubit from the
other Bell pair results in the three-qubit linear cluster:
\begin{equation}
\Qcircuit[2em] @R=1em @C=1em {
\node{} & \node{} \link{0}{-1} & \node{} \link{0}{-1}}
\end{equation}
If this fails we can try again from scratch.  Using this procedure, we
can obtain a supply of three-qubit linear clusters, with the average
cost of preparation a small constant.  Using such three-qubit linear
clusters as a resource, we can build up longer linear chains by
attempting to fuse three-qubit linear clusters onto the end of an
existing chain.  With probability $50$ percent this succeeds, adding
two qubits to the chain.  With probability $50$ percent it fails,
resulting in the loss of a qubit from the chain.  Thus, on average one
qubit is added to the chain, and standard results about random walks
imply that it is possible to build up a long linear chain with only a
small constant overhead.

More complex clusters can be built up using similar ideas.  To achieve
good thresholds, it's important to design the best possible procedures
for building up clusters.  This is a complex task, trading off two
opposing desiderata: (1) the need to keep the cluster generation
near-deterministic, which is most easily accomplished by preparing
large numbers of cluster qubits well in advance of when they are
measured (c.f.~\cite{Nielsen04b}), and (2) the fact that qubits left
to themselves tend to decay due to noise.  As a result, the exact
procedure we use to build up clusters in our fault-tolerant protocol
is rather involved, and we defer a detailed discussion until later in
the paper.

\section{Physical setting}
\label{sec:physical_setting}

In this section we describe in detail both what physical operations we
assume can be done, and our model of noise.

\textbf{Physical operations:} We assume the following basic elements
are available.  First, a source of polarization-entangled Bell pairs
(specifically, the state $[|0\rangle\otimes(|0\rangle+|1\rangle) + | 1
\rangle\otimes (|0\rangle-|1\rangle)]/2$ in qubit notation).
Physically, these can be produced in a number of different ways, but
the details don't matter to our analysis. Second, single-qubit gates
can be performed on the optical qubits. Physically, this can be done
using linear optics, following KLM. Third, the fusion gate of Browne
and Rudolph can be applied. Fourth, efficient
polarization-discriminating photon counters capable of distinguishing
$0, 1$ and $2$ photons are available.  These can be used to effect
measurements in the computational basis, and are also used to verify
the success or failure of the fusion gate.  Note that having
single-qubit gates and computational basis measurements allows us to
effect single-qubit measurements in an arbitrary basis. Single-qubit
gates do not appear explicitly in our protocol, rather only as part of
single-qubit measurements.

Implicit in our discussion up to now is the concept of a \emph{time
  step}.  For simplicity, assume that all the basic elements take the
same amount of time, and we describe our circuit in terms of a
sequence of such time steps.  As a consequence, an important
additional element that must be available is the quantum memory
``gate'', during which an optical qubit ideally does nothing for a
time step, but may still be affected by noise.  Physically, it's not
yet clear what the best way of implementing such a quantum memory will
be.

We've described the basic elements in our model of quantum
computation.  However, a number of important additional assumptions
are made about how these elements can be put together.  First, we
assume that any two qubits can be interacted directly.  This is
reasonable, given the ease of moving photonic qubits from one location
to another.  Second, we assume the ability to perform operations on
all the qubits in parallel.  Third, we assume the availability of
classical computation, communication, and feed-forward, all on a
timescale fast compared with the unit time step.  The feed-forward
requirement is particularly demanding, since it requires us to decide
which qubits interact in a time-step, based on the results of
measurements in the previous time-step.  To some extent, this
requirement is imposed merely to simplify our simulations, and it
seems likely that the requirement can be relaxed, but this remains a
topic for further investigation.

\textbf{Noise model:} We now describe our model of the physical
sources of noise. As stated in the introduction, our protocol deals
with three kinds of noise: (a) the inherent nondeterminism of the
fusion gates; (b) photon loss; and (c) depolarizing noise. We now
describe these in more detail.

In the last section we already described the noise due to the inherent
non-determinism of the fusion gate: with probability $50\%$ the gate
succeeds, and fusion is effected, while with probability $50\%$ it
fails, and the two qubits are measured in the computational basis.

We assume a single parameter $\gamma$ controls the strength of the
photon loss.  $\gamma$ is the probability per qubit per time step of a
photon being lost.  We assume this probability is independent of the
state of the qubit, and that photon loss affects every basic operation
in our protocol, as follows:

\begin{itemize}
\item Bell-state preparation: After the state has been prepared, each
  of the two qubits independently experiences photon loss with
  probability $\gamma$.

\item Memory, single-qubit, and fusion gates: Before the gate each
  input qubit experiences photon loss with probability $\gamma$.  In
  the case of the fusion gate, which has two inputs, we assume the
  loss probabilities are independent.  Later in the paper we also
  investigate the case where the photon loss rate for memory gates has
  been suppressed relative to the other gates.

\item Measurement: Before measurement we assume photon loss occurs
  with probability $\gamma$.  Physically, this can model both the loss
  of photons from the relevant optical modes, and also detector
  inefficiencies.

\end{itemize}

It is worth noting that detector inefficiencies are currently much
worse than other sources of photon loss, and it could be argued that
detector inefficiency and other photon loss rates should be treated as
independent parameters (or, alternately, that all other photon loss
rates be set to zero).  However, it is clear that turning off or
turning down photon loss noise in locations other than before
measurement can only improve the threshold, and so we have used the
more pessimistic model described above.  In fact, it can be shown that
photon loss occurring in locations other than before measurements
propagates to become equivalent to photon loss before measurements.
Thus, the model in which photon loss occurs only before measurement
should have a threshold for photon loss noise several times higher
than the results we report in this paper.

Note also that we have chosen a model of photon loss during Bell-state
preparation that acts independently on each qubit in the pair. It
would perhaps be more physically realistic for loss to occur in a
manner that is highly correlated between the two qubits in the Bell
pair (i.e., making it more likely that both photons in the pair are
lost as opposed to just one). However, the design of our
fault-tolerant protocol ensures that we can always detect situations
where both photons in a Bell state are lost, and thus this type of
coincidental loss has no negative effect on the threshold. So, our
choice of uncorrelated photon loss is the more pessimistic of the two
alternatives.

Similarly to photon loss, we assume a single \emph{depolarizing
  parameter} $\epsilon$ controls the strength of the depolarizing
noise.  We assume depolarization affects every basic operation, as
follows:

\begin{itemize}

\item Bell-state preparation: After the state has been prepared, the
  joint state of the two qubits is depolarized as follows: with
  probability $1-\epsilon$ nothing happens, while with respective
  probabilities $\epsilon/15$ we apply each of the $15$ non-identity
  Pauli product operators $IX, XX$ etcetera.

\item Memory and single-qubit gates: Before each gate we depolarize as
  follows: with probability $1-\epsilon$ nothing happens, while with
  respective probabilities $\epsilon/3$ we apply each of the $3$
  non-identity Pauli operators $X, Y$ and $Z$.

\item Fusion gate: The joint state of the two qubits is
  depolarized with parameter $\epsilon$ (in the same way as described for Bell-state preparation above) before being input to the
  gate.

\item Measurement: Before measurement the qubit is depolarized with
  parameter $\epsilon$ (in the same way as described for memory and single-qubit gates above).

\end{itemize}

Note that in our noise model noise occurs before or after operations.
In a real physical setting, noise will also occur during gate
operations.  However, standard fault-tolerance techniques (see,
e.g.,~\cite{Knill98a}) can be used to show that noise during an
operation can be regarded as completely equivalent to noise before or
after that operation.

The noise model we have described is obviously an approximation to
reality, and is incomplete in various ways.  For example, it is
difficult to justify on physical grounds using the same two noise
strength parameters for all operation types. Also, additional noise
sources that may have an effect in real implementations include dark
counts, dephasing, and non-local correlations.  However, it can be
shown that the fault-tolerant protocol we implement automatically
provides protection against such noise sources.  We haven't done a
detailed investigation of the threshold for these noise sources, or
for the case of different noise strengths for different operation
types, but believe that the results would be in qualitative agreement
with the results of the present paper.

\section{How we simulate a noisy cluster-state computation}
\label{sec:simulations}

In this section we explain how to simulate a noisy cluster-state
computation.  We don't yet describe the details of the fault-tolerant
protocol, leaving those to the next section.  However, the protocol is
simulated using essentially the techniques we now describe.

We concentrate on the case when the errors are solely Pauli-type
errors.  It turns out that a simple modification of the techniques
used to describe these errors can be used to describe the
non-deterministic failure of fusion gates, or photon loss.  However,
we defer this discussion to the next section as it depends on some
details of the fault-tolerant protocol.

\subsection{Example}
\label{subsec:example_noisy_csc}

We begin with a toy example of a noisy cluster-state computation in
which noise is introduced at just a single location, and we study how
this affects the remainder of the computation.  This example will
motivate our later abstractions and the data structures used to model
noise.

The example is a two-qubit cluster-state computation:
\begin{equation}
\Qcircuit[4em] @R=1em @C=1em {
 \node{H} & \node{} \link{0}{-1}
}
\end{equation}
We imagine that the two qubits of the cluster are perfectly prepared.
After preparation, we suppose a single Pauli $Z$ error corrupts the
first qubit, so the actual physical state of the cluster is related to
the ideal state by an overall error $Z\otimes I$. Now we suppose a
perfect $H$ operation and computational basis measurement is carried
out on the first qubit, yielding an outcome $m =0$ or $1$.  It will be
convenient to regard the combined Hadamard and measurement as a single
operation, a perfect measurement in the $X$ basis.  This completes our
example computation.

At the end of the computation, the experimenter believes the resulting
state of the second qubit is $X^m H |+\rangle$.  However, a
calculation shows that the actual state is $X^{m+1} H |+\rangle$.
Mathematically, there are two different ways we can think about this
resulting state:
\begin{itemize}
\item Measurement of the first qubit propagates the $Z$ error on
that
  qubit to the second qubit, and causes it to become a physical $X$
  error on that qubit.

\item Measurement of the first qubit causes the $Z$ error on that
  qubit to turn into an $X$ error in the Pauli frame of the second
  qubit, but eliminates all physical errors.
\end{itemize}
While these points of view are equivalent, we will take the second
point of view, as it turns out that in more complex examples, it is
this point of view which gives the simplest description of what is
going on.

This analysis can be repeated for the case where, instead of a $Z$
error, we had a single $X$ error occur on the first qubit.  However,
this case is more trivial, because the $X$ error followed by the
perfect $X$ basis measurement is equivalent to a perfect $X$ basis
measurement alone, and thus the resulting state is $X^m H|+\rangle$,
as expected by the experimenter.  Thus, in this case the effect of
measurement is simply to eliminate the physical error.

\subsection{General description of the introduction and propagation
  of Pauli noise}

Our simulations of Pauli noise in cluster-state computation are based
on generalizations of the concepts introduced in the previous example.
There are two basic data structures that we keep track of.  The first
is the \emph{physical error} in the state of the cluster.  This is a
tensor product of Pauli matrices, one for each cluster qubit.  This
begins as the identity at every qubit, and we will describe below how
it is modified as noise and gate operations occur.

The second data structure is the \emph{error in the Pauli frame} of
the cluster.  Again, this is a tensor product of Pauli matrices, one
for each qubit in the cluster.  It begins as the identity at every
qubit, and will be modified during the simulation according to rules
described below.

It is notable that our description of noisy cluster-state computation
is thus based entirely on products of Pauli operators. What makes this
description possible is that all the operations we simulate are
Clifford-group operations, and this ensures that the errors remain
Pauli errors at all times.  It is also worth noting that in our
simulations we do not keep track of the actual state of the cluster,
nor of the Pauli frame, but only of the errors in each.  This is
because the aim of our fault-tolerance simulations is to determine
various statistics associated to these errors, and the actual state of
the cluster is not of direct importance.

Note that in our description, physical errors and Pauli frame errors
are not generally interchangeable, since they undergo different
propagation rules (described later in this subsection) and thus may
have different effects on the final state of the computation. Errors
in the Pauli frame are introduced only as a result of noise-affected
measurements in transport circuits, and propagate as a result of the
Pauli frame update rules (described in Subsection
\ref{subsec:cluster_states}) that the experimenter applies. Physical
errors describe noise on the state itself, and propagate according to
how the Pauli matrices commute through the various quantum operations
performed on the state.

As we have described, the physical error and Pauli frame error are
products of Pauli operators on all the remaining cluster qubits.
It is often convenient to focus on one or just a few cluster
qubits rather than the entirety.  For this purpose we will refer
to \emph{local physical errors} and \emph{local Pauli frame
errors}, which are just the corresponding Pauli operators for a
specified qubit or qubits.  It will also be convenient to describe
such local errors either in matrix form as $X^xZ^z$, or in terms
of the pair $(x,z)$, and we will use these descriptions
interchangeably.  So, for example, we may refer to either $X^x$ or
simply $x$ as the $X$ error.  We will routinely ignore global
phase factors in our description of errors, so, e.g., $XZ$ and
$ZX$ are regarded as equivalent.

The final concept needed to explain the way we update our data
structures is that of a \emph{terminating qubit}.  We define a cluster
qubit to be terminating if it has no horizontal bonds.  It may or may
not have vertical bonds.  For example, in a horizontal cluster being
used to simulate a single-qubit computation, the qubit at the
rightmost end of the cluster becomes a terminating qubit after all the
other qubits have been measured.  The significance of terminating
qubits in a cluster-state computation is that measurement of these
qubits reveals the outcomes of the computation.  By contrast,
measurement of non-terminating qubits merely reveals information which
can be used to propagate quantum information to other parts of the
cluster.

We now describe the rules for updating both our data structures for
each of the possible operations that can occur during a noisy
cluster-state computation.

\emph{Update rule for depolarization event:} The physical error is
updated by matrix multiplication by the appropriate randomly-chosen
error (e.g., $X, Y$ or $Z$).  The error in the Pauli frame is not
affected.

\emph{Update rule when a non-terminating qubit is measured in the $X$
  basis:} It is easiest to describe this by describing two separate
cases: the case when there is a single horizontal bond attached to the
qubit being measured, to the right; and the case where both vertical
and horizontal bonds are present.

Suppose the qubit being measured has a single horizontal bond
attached, to the right.  Suppose before the measurement the local
physical error on the qubit being measured is $X^{x_{1p}} Z^{z_{1p}}$,
and the local Pauli frame error is $X^{x_{1f}} Z^{z_{1f}}$.  Suppose
the corresponding values for the qubit on the right are $X^{x_{2p}}
Z^{z_{2p}}$ and $X^{x_{2f}} Z^{z_{2f}}$.  After the measurement the
updated values for the local physical and Pauli frame errors of the
qubit on the right are as follows:
\begin{eqnarray}
  x_{2p}' & = & x_{2p} \label{eq:horizontal_error_update_1} \\
  z_{2p}' & = & z_{2p} \label{eq:horizontal_error_update_2} \\
  x_{2f}' & = & x_{2f}+z_{1p}+z_{1f} \label{eq:horizontal_error_update_3} \\
  z_{2f}' & = & z_{2f}+x_{1f}. \label{eq:horizontal_error_update_4}
\end{eqnarray}
These rules are derived from our earlier description of the transport
circuit and the rules for updating the Pauli frame, along essentially
the same lines as the example in
Subsec.~\ref{subsec:example_noisy_csc}.  As in the example, we see
that $X$ physical errors on the qubit being measured are eliminated,
and $Z$ physical errors propagate to become $X$ errors in the Pauli
frame.  Similar reasoning shows that $X$ errors in the Pauli frame of
the qubit being measured propagate to become $Z$ errors in the Pauli
frame of the attached qubit, and vice versa for $Z$ errors in the
Pauli frame.

Suppose the qubit being measured has a vertical bond, and a rightward
horizontal bond.  Suppose we label the qubits $1$ (qubit being
measured), $2$ (qubit to the right), and $3$ (qubit attached by
vertical bond).  We will denote the values for the local physical
error and local Pauli frame error by $x_{jp}, z_{jp}$ and $x_{jp},
z_{jp}$, respectively, where $j$ labels the qubit.  We update these in
two stages, with the update method derived from the two stages for
updating the Pauli frame when a vertical bond is present, as described
in Subsec.~\ref{subsec:cluster_states}.  The first stage is associated
to the vertical bond.  We set:
\begin{eqnarray}
  x_{1p}' & = & x_{1p} \label{eq:vertical_error_update_1} \\
  z_{1p}' & = & z_{1p} \label{eq:vertical_error_update_2} \\
  x_{1f}' & = & x_{1f} \label{eq:vertical_error_update_3} \\
  z_{1f}' & = & z_{1f}+x_{3f} \label{eq:vertical_error_update_4} \\
  x_{3p}' & = & x_{3p} \label{eq:vertical_error_update_5} \\
  z_{3p}' & = & z_{3p} \label{eq:vertical_error_update_6} \\
  x_{3f}' & = & x_{3f} \label{eq:vertical_error_update_7} \\
  z_{3f}' & = & z_{3f}+x_{1f}. \label{eq:vertical_error_update_8}
\end{eqnarray}
The local physical and Pauli frame errors for qubit $2$ are not
changed during this step.  For the second stage we behave as though
the vertical bond has been deleted, and use our new values for the
physical and Pauli frame errors as input to the update rules for the
case of a horizontal bond,
Equations~(\ref{eq:horizontal_error_update_1})-(\ref{eq:horizontal_error_update_4}).

\emph{Update rule for measuring terminating qubits in the $X$ basis:}
We first describe the update rules for the case when the terminating
qubit has no vertical bonds attached.  The update rule is to compute
the \emph{total error}, which we define as the product of the local
physical and Pauli frame errors on that qubit.  The qubit is then
deleted from the cluster, and its local physical and Pauli frame
errors are deleted from the corresponding data structures.  The total
error $X^x Z^z$ determines whether or not the measurement outcome
(e.g., of syndrome information) contains an error.  Since the
measurement is in the $X$ basis, the error in the measurement is
simply $z$.  The aim of our fault-tolerant simulations will be to
determine various statistics associated to this total error.

Consider now the case when the terminating qubit has a vertical bond
also, before being measured in the $X$ basis.  In this case we simply
follow the rules of
Equations~(\ref{eq:vertical_error_update_1})-(\ref{eq:vertical_error_update_8})
for updating the errors, and then treat the qubit as though the
vertical bond has been deleted, and apply the rules described earlier
for treating a terminating qubit.

\emph{Update rule for measuring a qubit in the $Z$ basis:} In our
protocols we only ever do such a measurement on non-terminating
qubits, and so restrict our attention to this case.  In an ideal
cluster-state computation the effect of a $Z$ measurement with outcome
$m = 0$ or $1$ is effectively to remove that qubit from the
cluster~\cite{Raussendorf01a}, and apply $Z^m$ to all neighbouring
qubits.  An experimenter getting a result $m$ can therefore update the
Pauli frame of neighbouring qubits by multiplying each by an extra
factor of $Z^m$.

To describe the update rule in this case, we define the total error to
be $x_t = x_p+x_f, z_t = z_p+z_f$, where subscript $p$s denote
physical errors, and subscript $f$s denote Pauli frame errors.  The
error in the measurement outcome will be $x_t$, since $Z$ flips don't
affect $Z$ basis measurements.  So the update rule is merely to
discard the local physical and Pauli frame errors from our overall
physical error and Pauli frame error, and to introduce an additional
$Z^{x_t}$ Pauli frame error on all neighbouring qubits.

\emph{Update rules for the fusion gate:} We separate our analysis into
cases when the fusion gate is unsuccessful and successful.  When
unsuccessful the fusion gate results in a $Z$ basis measurement being
applied to the qubits we are attempting to fuse.  This case can be
described by the rules stated above for $Z$ basis measurements.

When the fusion gate is successful we update as follows. We label the
qubits being fused as qubit $1$ and $2$.  It turns out that in our
fault-tolerant protocol we \emph{never} fuse qubits which have Pauli
frame errors.  Thus we can assume that the initial errors on the
qubits being fused are simply $x_{jp}, z_{jp}$, where $j=1, 2$ labels
the qubit.  For distinctness we will call the physical and Pauli frame
errors of the output qubit $x_{3p}$ and $z_{3p}$; the $3$ is merely
for clarity, and does not indicate the creation of a new physical
qubit.  The update rule is as follows:
\begin{itemize}
\item For each qubit neighbouring qubit $1$, we add $x_{2p}$ to the
  $Z$ physical error.

\item Vice versa, for each qubit neighbouring qubit $2$, we add
  $x_{1p}$ to the $Z$ physical error.

\item $x_{3p} = x_{1p}+x_{2p}$,

\item $z_{3p} = z_{1p}+z_{2p}$.

\end{itemize}
These rules follow straightforwardly from the definition of the fusion
gate.

\section{Fault-tolerant protocol}
\label{sec:protocol}

\subsection{Introduction}

In this section we describe in detail our fault-tolerant protocol, and
the threshold results we obtain.  We begin in this subsection with a
brief discussion of the historical background and antecedents to our
work.  We begin describing the technical details of the protocol in
the next subsection.

The basic theory of fault-tolerant error-correction originated in the
mid-1990s, with the fault-tolerant constructions of
Shor~\cite{Shor96b}.  These constructions were used to prove a
threshold theorem for quantum computation by Aharonov and
Ben-Or~\cite{Aharonov97a,Aharonov99a}, Gottesman and Preskill (see,
e.g.,~\cite{Aliferis05b,Reichardt05a,Gottesman98a,Gottesman97a,Preskill98b}),
Kitaev~\cite{Kitaev97a}, and Knill, Laflamme and
Zurek~\cite{Knill98a}.  This work established the \emph{existence} of
a threshold for a wide class of noise models, and gave pessimistic
analytic bounds on the threshold, but did not establish the exact
value of the threshold.

A large body of numerical work aimed at determining the threshold has
since been done.  Especially notable is the work by
Steane~\cite{Steane03a}, who did the first detailed numerical
investigations of the threshold, and the recent work by
Knill~\cite{Knill05a}, who has established the best known thresholds
in the standard quantum circuit model.  Many of our techniques are
based on those described by Steane.  We will also see below that there
is some overlap with the techniques of Knill.  Of course, the very
different nature of optical cluster-state computation demands many new
techniques, and care must be taken in comparing the values of
thresholds in this model and the standard quantum circuit model.

As explained in the introduction, Nielsen and Dawson~\cite{Nielsen05a}
proved the existence of a threshold for optical cluster-state
computing (c.f.~\cite{Raussendorf03b,Aliferis05a,Raussendorf05a}).
The basic idea of the construction in~\cite{Nielsen05a} is to show
that if we take a quantum circuit, convert it into a fault-tolerant
form using multiple layers of concatenated coding, and then simulate
the circuit using optical cluster states, the resulting noisy optical
cluster-state computation is itself fault-tolerant.  This proof relied
on an important theorem of Terhal and Burkard~\cite{Terhal04a}
establishing a threshold for non-Markovian noise in the standard
circuit model.

When we began the work described in this paper, our intention was to
apply the procedure described in~\cite{Nielsen05a} to a fault-tolerant
circuit protocol similar to that considered by
Steane~\cite{Steane03a}.  In the event, our protocol involves
substantial improvements over this basic procedure, is manifestly
fault-tolerant, and gives a much better threshold.  The improvements
include: optimizing our treatment of photon loss and non-determinism;
exploiting the ability to premeasure and parallelize parts of the
cluster-state computation; and taking advantage of the ability to
premeasure clusters in order to improve ancilla creation.  All these
improvements are described in detail below.

\subsection{Broad picture of fault-tolerant protocol}

In this subsection we outline our fault-tolerant protocol.  The
protocol is split into two main parts.

The first part is a cluster-based simulation of a variant of Steane's
fault-tolerant protocol. We have modified Steane's protocol to deal
with the non-deterministic nature of the optical gates, and introduced
several cluster-based tricks to improve the threshold.  This protocol
and the results of our simulations are described in
Subsection~\ref{subsec:ft_protocol_cluster}.

The second part is a deterministic gate-based protocol, whose purpose
will be explained in the paragraphs below.  This protocol is also
based on Steane's methods, again with some substantial variations.
This protocol and the results of our simulations are described in
Subsection~\ref{subsec:ft_protocol_deterministic}.

The reason for using the two protocols is that the actual cluster
threshold is obtained by concatenating a single encoded level of the
cluster protocol with multiple levels of the deterministic protocol.
This works because our multiply concatenated fault-tolerant cluster
protocol is equivalent to building up a fault-tolerant implementation
through multiple levels of concatenation in the circuit model, and
then replacing each gate in the bottom level by a clusterized
equivalent.

To obtain the overall behaviour of such a protocol, it is not feasible
to directly simulate the multiply concatenated computation.  Instead,
we do one simulation of the clusterized protocol at just a single
level of encoding, and another of the deterministic protocol, again at
a single level of encoding.  We then make an argument allowing us to
use the data obtained from these two protocols to estimate the overall
behaviour if multiple layers of concatenation had in fact been used.
The details of how this is done are described in
Subsection~\ref{subsec:results}.

\subsection{The cluster-based protocol}
\label{subsec:ft_protocol_cluster}

Our cluster-based protocol performs multiple rounds of clusterized
quantum error-correction, effectively implementing a fault-tolerant
quantum memory.  Following previous numerical work on the threshold
(e.g.~\cite{Steane03a,Knill05a}) we do not simulate dynamical
operations at the encoded level.  However, our simulations could
easily be varied to implement encoded Clifford group operations with a
small additional overhead, and this will leave the threshold
essentially unchanged. Computational universality requires at least
one encoded non-Clifford group operation.  This is difficult to
simulate, and previous workers~\cite{Steane03a} have argued that it
changes the threshold very little, since error-correction (which makes
up the bulk of a fault-tolerant circuit) is done using only Clifford
group operations.

Our simulations extract various statistics regarding failure modes
of our fault-tolerant protocol.  Thus we do multiple trials of the
protocol in order to estimate these statistics.  A single trial
involves the simulation of multiple rounds of quantum
error-correction applied to a single encoded logical qubit.  This
is all done within the optical cluster-state model of computation,
with the noise model as described in
Section~\ref{sec:physical_setting}.

The major elements of a single trial are as follows: (1) the input
state; (2) the input to a round of quantum error-correction; (3) the
preparation of the ancilla states used to extract error syndromes; (4)
the preparation and use of the \emph{telecorrector} cluster enabling
interactions between the encoded data and the ancilla; (5) the
reduction of photon loss and non-determinism to Pauli-type errors; and
(6) decoding.

We will now describe each of these elements in detail.  First,
however, we discuss some special tools which are used repeatedly in
multiple elements of our cluster-state computation.

\subsubsection{Tools for optical cluster-state computing: microclusters,
  parallel fusion, and postselection}

Earlier in the paper we've described how to clusterize quantum
circuits, and how to implement cluster-state computation
optically. However there are three useful additional tools which
we use repeatedly through the entire protocol, and which deserve
special mention: \emph{microclusters}, \emph{parallel fusion}, and
\emph{postselection}.

A microcluster is a star-shaped cluster, for example:
\begin{equation}
\Qcircuit[2em] @R=1em @C=1em {
        & \node{} \link{1}{0}  & \\
        & \node{} \link{1}{-1}  & \node{} \link{0}{-1} \\
\node{} &                      & \node{} \link{-1}{-1}
}
\end{equation}
The central node in the microcluster is known as the \emph{root node},
while the other nodes are \emph{leaf nodes}.  Microclusters are used
as a tool to build up larger clusters.  In particular, the use of
microclusters ensures that these larger clusters always have multiple
leaf nodes, which can be used to enhance the probability with which we
fuse two larger clusters:
\begin{equation}
\Qcircuit[1em] @R=1em @C=1em {
        &                      &                      &
& \node{} \link{1}{-1} & \node{} \link{1}{1} \\
\node{} & \node{} \link{0}{-1} & \node{} \link{0}{-1} & \node{} \link{0}{-1}
& \node{} \link{0}{-1} & \node{}            & \node{} \link{0}{-1} & \node{} \link{0}{-1} & \node{} \link{0}{-1} & \node{} \link{0}{-1} \\
        &                      &                      &
& \node{} \link{-1}{-1} & \node{} \link{-1}{1}
}
\end{equation}
We can attempt three simultaneous fusion gates between adjacent leaf
nodes of the two clusters.  With a probability that goes rapidly to
one as the number of leaves increases, at least one of these fusion
gates succeeds, fusing the two clusters:
\begin{equation}
\Qcircuit[1em] @R=1em @C=1em {
\node{} & \node{} \link{0}{-1} & \node{} \link{0}{-1} & \node{} \link{0}{-1}
& \node{} \link{0}{-1} & \node{} \link{0}{-1} & \node{} \link{0}{-1} & \node{} \link{0}{-1} & \node{} \link{0}{-1}
}
\end{equation}
When more than one fusion gate succeeds, we can obtain the same fused
cluster, simply by measuring redundant fused nodes in the
computational basis\footnote{In fact, the
  operations we need to do can even be accomplished without removing
  any redundant nodes, and this is the approach we take in our
  simulations. In particular, imagine that $k$ of the simultaneous
  fusions succeed, resulting in $k$ qubits in a position where there
  should be just one. It can be shown that this cluster state is
  stabilized by (that is, is a $+1$ eigenstate of) a tensor product of
  $X$s on any even number of those $k$ qubits. This shows that if we
  were to later measure one of the $k$ qubits in the $X$ basis, as
  part of the normal running of the cluster, then the state of each of
  the other $k-1$ extra qubits would collapse to an eigenstate of $X$,
  thus automatically disentangling them from the cluster without the
  need for further measurements. Note that there is a potential
  advantage in measuring the extra qubits anyway, in the $X$ basis, to
  verify the measurement outcome of the first qubit. However, we do
  not perform this type of verification in the simulations.
  }.  We call this process of using leaves to fuse the two
clusters with high probability \emph{parallel fusion}.

We try to create microclusters in a way that meets two complementary
aims: (1) we wish to create them rapidly, in order to minimize the
effects of noise; and (2) we wish to use the fewest physical resources
possible in creating the microclusters.  Our microcluster creation
protocol is designed with both these goals in mind; somewhat better
thresholds could be obtained at the expense of using more resources.

When the number of leaves is a power of two, e.g., $k = 2^m$, we
create the microcluster as follows.  We begin with $2^m$ one-leaf
microclusters, which are just Bell pairs:
\begin{equation}
\Qcircuit[2em] @R=1em @C=1em {
\node{} \link{1}{0}  & \node{} \link{1}{0} & \node{} \link{1}{0} & \node{} \link{1}{0} \\
\node{}              & \node{}             & \node{}             & \node{}
}
\end{equation}
We then fuse pairs of the one-leaf microclusters in order to create
two-leaf microclusters:
\begin{equation}
\Qcircuit[2em] @R=1em @C=1em {
\node{} \link{1}{1}  & \node{} \link{1}{0} & \node{} \link{1}{1} & \node{} \link{1}{0} \\
                     & \node{}             &                     & \node{}
}
\end{equation}
We continue in this way, repeatedly fusing the root nodes of pairs of
microclusters, obtaining microclusters with ever more leaves.  For the
$4$-leaf case, the process terminates at the next stage:
\begin{equation}
\Qcircuit[2em] @R=1em @C=1em {
\node{} \link{1}{1}  & \node{} \link{1}{0} & \node{} \link{1}{-1} & \node{} \link{1}{-2} \\
                     & \node{}             &                     &
}
\end{equation}
The protocol when the number of leaves is not a power of two is a
straightforward modification.

When preparing the microclusters, the fusion gates will inevitably
sometimes fail.  However, by doing a large number of attempts to
create the microcluster in parallel, we can ensure that with very high
probability at least one of these attempts will be successful.  For
simplicity, in our simulations we assume that fusion gates are
\emph{always} successful during microcluster creation (but not in
general).  This is justified because the experimenter can always
postselect during microcluster creation. With this postselection, the
expected number of Bell pairs consumed per $k$-leaf microcluster is
$k^2$, and it takes $\log_2(k)+1$ time-steps to create the
microcluster.

Our use of postselection in microcluster creation is merely one place
at which we use postselection.  It can be used whenever performing
manipulations on clusters that do not contain any of the data being
processed.  This will include ancilla and telecorrector creation,
which actually contain the bulk of the operations performed in our
computation.  This is extremely convenient, for it enables us to
assume that non-deterministic operations have been performed
successfully, at the expense of requiring the experimenter to perform
a number of attempts at such operations in parallel, and to postselect
on the successful operations.  It will be important for us to keep
track of the scaling involved in such postselection, to ensure that no
exponential overheads are incurred.

\subsubsection{Input states}

The trials we simulate consist of multiple rounds of clusterized
quantum error-correction.  To describe how these rounds occur we must
first specify the form of the state which is input to a round.  The
first round of error-correction is, of course, somewhat special, since
it's the initial state of the entire computation.  Nonetheless, it has
the same general form as the inputs to any other round.  Therefore, we
begin by describing the general case, before discussing some caveats
specific to the initial state of the entire trial.

The state of our optical cluster-state computer at the start of any
given round is of the following form:
\begin{equation} \label{eq:input_state_to_round}
\Qcircuit[0.6em] @R=0.1em @C=1em {
        & \node{} \link{1}{-1} \\
\otimes & \node{} \link{0}{-1} \\
        & \node{} \link{-1}{-1} \\
        & \node{} \link{1}{-1} \\
\otimes & \node{} \link{0}{-1} \\
        & \node{} \link{-1}{-1} \\
        & \node{} \link{1}{-1} \\
\otimes & \node{} \link{0}{-1} \\
        & \node{} \link{-1}{-1} \\
        & \node{} \link{1}{-1} \\
\otimes & \node{} \link{0}{-1} \\
        & \node{} \link{-1}{-1} \\
        & \node{} \link{1}{-1} \\
\otimes & \node{} \link{0}{-1} \\
        & \node{} \link{-1}{-1} \\
        & \node{} \link{1}{-1} \\
\otimes & \node{} \link{0}{-1} \\
        & \node{} \link{-1}{-1} \\
        & \node{} \link{1}{-1} \\
\otimes & \node{} \link{0}{-1} \\
        & \node{} \link{-1}{-1}
\gategroup{2}{1}{20}{1}{1em}{--}
}
\end{equation}
This is not (quite) a cluster state.  To describe the state in the
ideal case, consider the following two-stage preparation
procedure\footnote{This is, of course, not the actual procedure used
  to obtain the state, but merely a convenient way of describing what
  the state is.}: (1) prepare the boxed qubits (i.e. the root nodes)
in the encoded state of the corresponding qubit; and (2) attach bonds
to the leaves according to the standard definition.  We will make use
of the leaves in the manner described earlier, to enhance the
probability of fusing this input cluster to the telecorrector state
(described later), which is used to effect the error-correction.  As
pictured, we have three leaves per root node, however in simulations
this number may be varied.

Of course, in practice, the actual state will be related to this ideal
state by a Pauli frame, and possibly also affected by noise in the
Pauli frame, and on the physical qubits.  These deviations are
described using the techniques we have already introduced.
Furthermore, in practice the root nodes will typically have been
premeasured, and so won't actually be physically present.  However, as
we have argued earlier, it is often convenient to carry out the
analysis as though operations were done in a different order than is
actually the case physically, and so we will sometimes describe the
computation as though the root nodes (and the associated local Pauli
frames) are present at the beginning of the round.

At the beginning of the entire trial, we assume the input is a
noise-free state of the form depicted in
Equation~(\ref{eq:input_state_to_round}).  Of course, in practice, the
actual state at the beginning of the computation will be noisy.
However, this noise-free assumption is justified on the grounds that
the initial state does not actually matter, since our goal is to
estimate the rate \emph{per round} at which crashes are introduced
into the encoded data.  Following Steane~\cite{Steane03a}, we perform
some number of ``warm-up'' rounds of error-correction before beginning
to gather data on this crash rate, in order to avoid transient effects
due to the particular choice of initial state.  The reason for
starting with a noise free state is because it is a reasonable
approximation to the actual (noisy) state of the computer after many
rounds, and thus the transient effects can be expected to die out
relatively quickly compared with many other possible starting states.

\subsubsection{Ancilla creation}
\label{sec:ancilla_creation}

Each round of quantum error-correction involves the creation of some
number of verified ancilla states, which are used to extract syndrome
bits.  These states are analogous to the ancillas used in standard
fault-tolerant quantum computation.  The exact number of ancillas
required may vary from round to round; we describe later the details
of how they are integrated into the computation.

In this section we describe the cluster-state computation used to
prepare a single ancilla.  This computation is essentially a
clusterized version of Steane's~\cite{Steane02a} ancilla creation
circuit.  We will describe this for the case of the Steane $7$-qubit
code, but the procedure generalizes in a straightforward manner to
many other Calderbank Shor Steane (CSS) codes, including the 23-qubit
Golay code used in some of our simulations\footnote{The 23-qubit Golay
code is derived from the classical binary Golay code, whose defining
parity check matrix is given in, for example, Sec.~5.3.3 of
\cite{Ling04a} and online at \cite{Weisstein06a}. }.

Following Steane, we can create an ancilla for the $7$-qubit code
using a quantum circuit of the form:
\begin{equation}
\epsfxsize=8.5cm \epsfbox{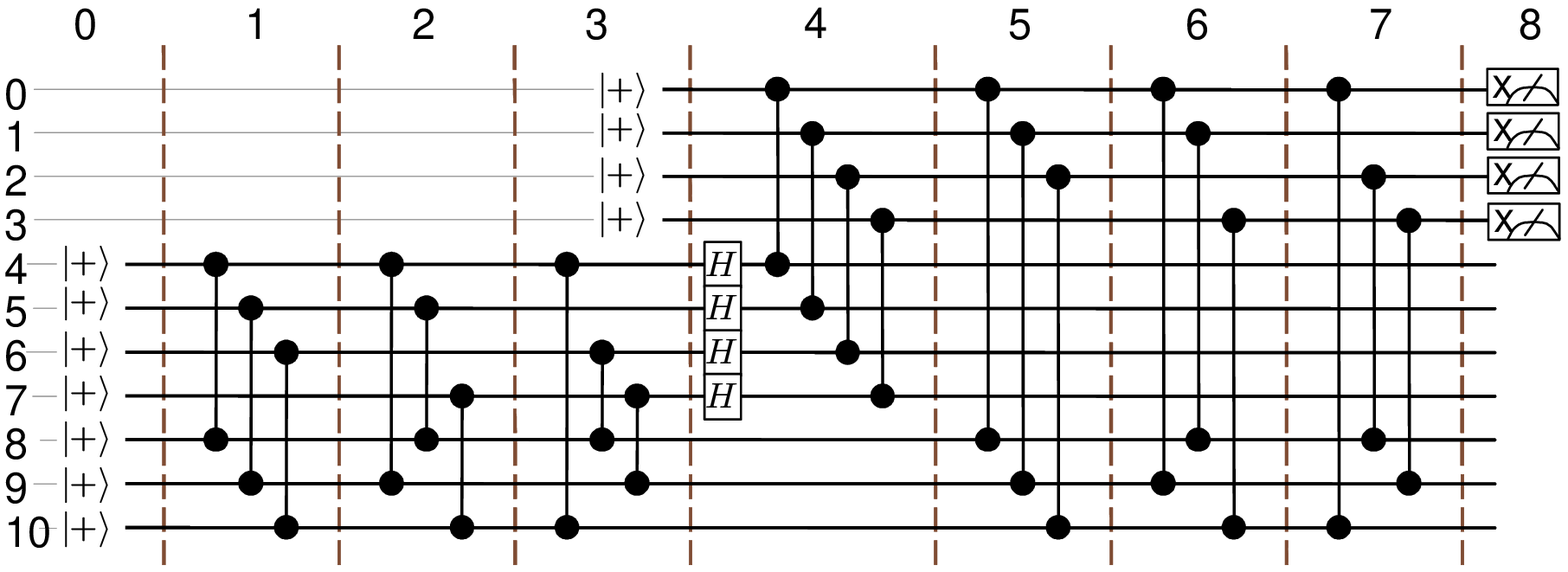}
\end{equation}
We clusterize this following the standard procedures (as described
in Section~\ref{subsec:cluster_states}) for clusterization, but
optimized in order to meet two complementary goals: (1) we do many
operations in parallel, in order to reduce the effects of noise;
and (2) careful use of postselection, in order to prevent a
blow out in resource usage.

We begin the clusterization by creating an array of microclusters:
\begin{equation}
\hspace{-1ex} \epsfxsize=8.7cm \epsfbox{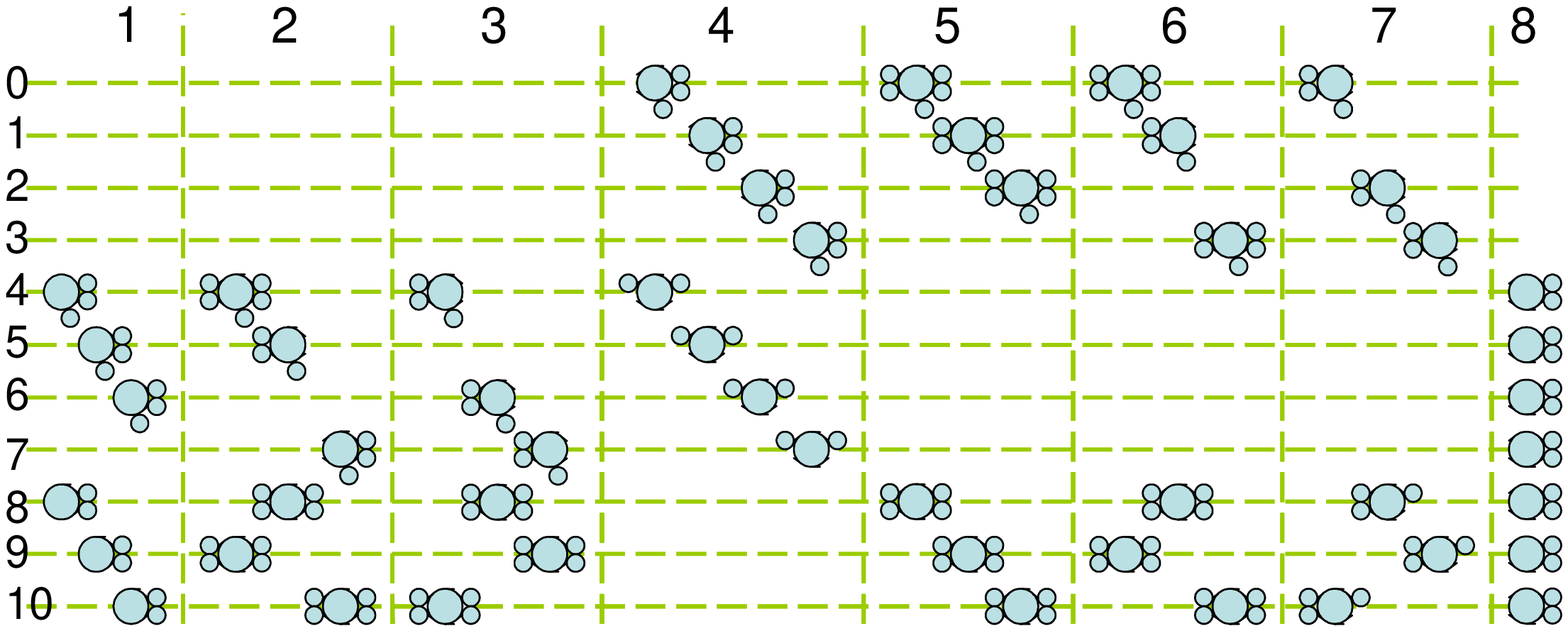}
\end{equation}
For clarity we have abridged our microcluster notation, omitting the
bonds, and just drawing the nodes; the large circles are root nodes,
while the small circles are leaf nodes.  The co-ordinates in our
microcluster array (denoted by the dashed lines and numbered labels)
correspond directly to the co-ordinates in the Steane circuit. The
only exception is the final column of the microcluster array, which
corresponds to the output of the cluster-state computation. Nontrivial
gates in the Steane circuit are replaced by microclusters, while
memory steps do not require additional microclusters, and so we omit
these where possible.

Our next goal is to create the following bonded microcluster
array:
\begin{equation}
\hspace{-1ex}\epsfxsize=8.7cm \epsfbox{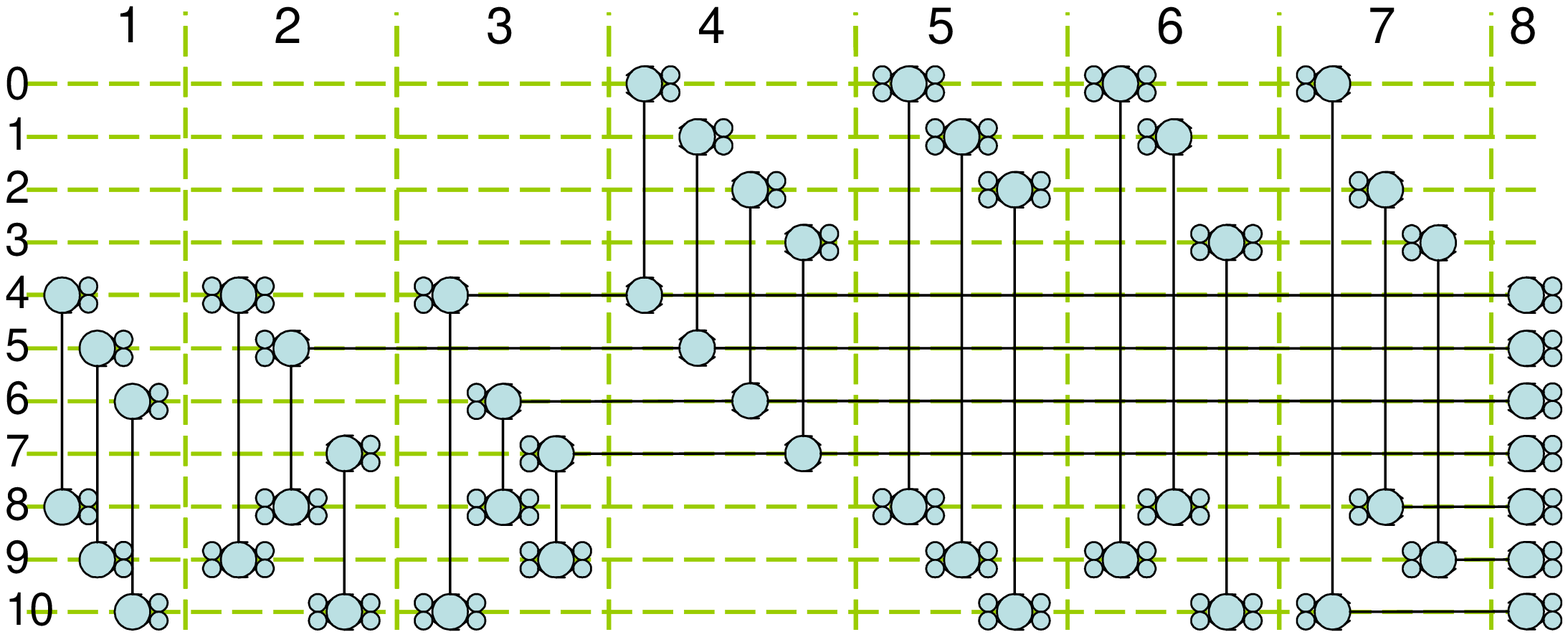}
\end{equation}
We do this in two steps.  The first step is to attempt creation of all
the \emph{vertical} bonds, by fusion of appropriate leaves and roots.
By postselection we can assume that all of these fusions were
successful and no photon loss was detected.  In reality, the
experimenter will need to create a larger array of microclusters, and
attempt all the fusions simultaneously, discarding wherever
unsuccessful.  The second step is to create the horizontal bonds,
again by fusions of the appropriate leaves and roots, and using
postselection to ensure success.

The final step is to obtain the cluster:
\begin{equation}
\epsfxsize=8.5cm \epsfbox{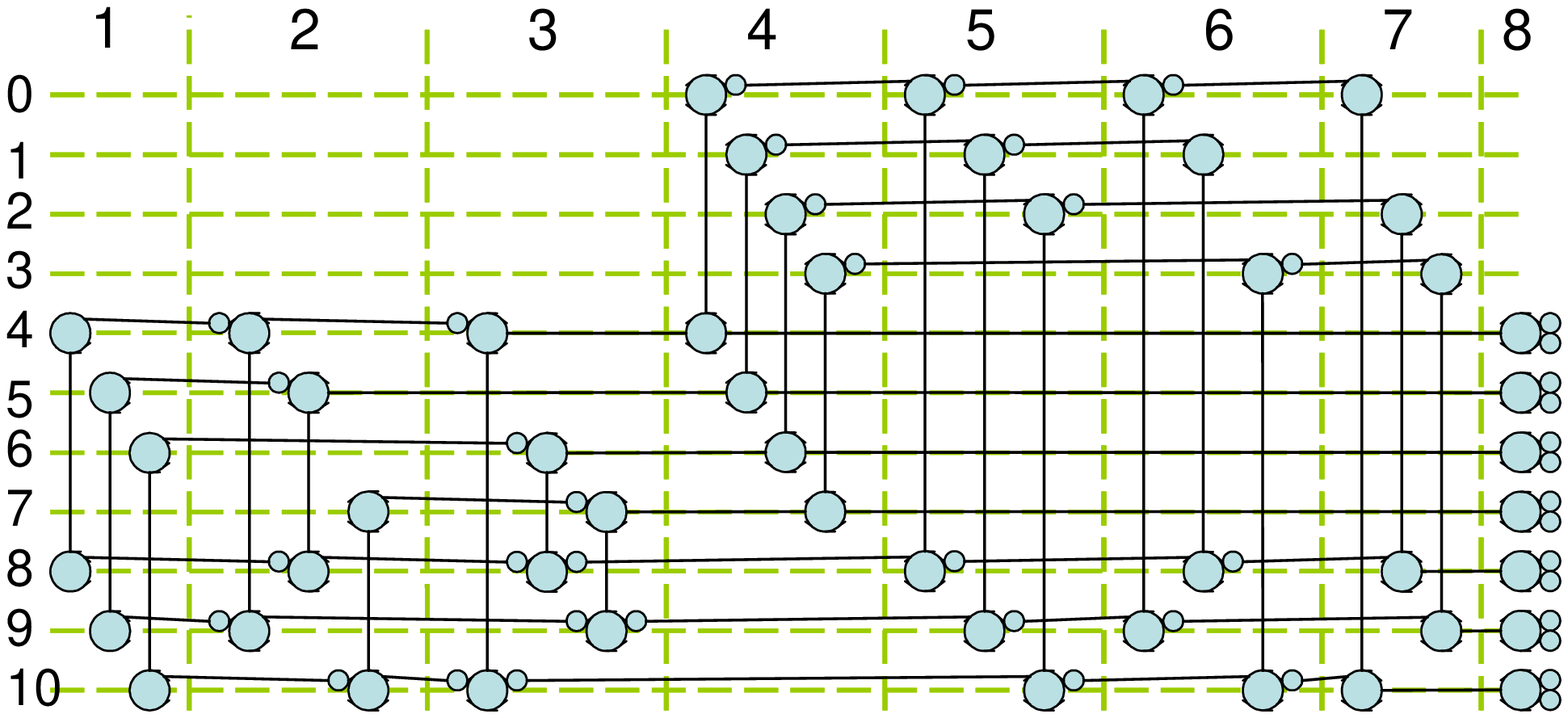}
\end{equation}
using parallel fusion to add the remaining horizontal bonds.  The
reason we use parallel fusion at this stage is to reduce the cost of
postselection.  The horizontal bonds added at this stage connect large
parts of the cluster, and so a failure of any will result in the need
to start over, and thus it is important to ensure a high probability
of success, in order to reduce resource usage.

Note that, as illustrated, parallel fusion involves $2$ attempted
connections.  However, the number of attempted connections is a
variable of our simulation, and in practice we have been using 3.
Varying this figure will affect both the noise threshold and the
resource usage. A value of $1$ is the best choice with respect to the
noise threshold, since the microclusters used would be smallest in
this instance, thus creating less opportunities for noise to be
introduced. The corresponding resource overhead would be particularly
bad though, due to the very small probability ($\frac{1}{2^{29}}
\approx 2\times10^{-9}$ for the $7$-qubit code) of fusion gates in the
final step of ancilla cluster creation all succeeding. Using $3$
attempts per parallel fusion, the probability of success of the final
step increases to $\frac{7^{29}}{8^{29}} \approx 0.02$. If the number
of attempts per parallel fusion is made too large, the benefit to the
resource usage due to the higher probability of success is outweighed
by the expense of creating large microclusters at the beginning. We
have not performed a detailed analysis of the optimal choice for this
parameter, rather we have chosen $3$ as a reasonable trade-off between
noise performance and resource usage.


To conclude the ancilla preparation, we simultaneously measure all
remaining qubits in the $X$ basis, except those qubits in column $8$,
applying the standard rules for Pauli frame propagation.  To verify
the ancilla, we postselect on the measurement results of the
terminating qubits in rows $0, 1, 2, 3$ all being $0$.  The resulting
state is identical to the state illustrated in another context in
Equation~(\ref{eq:input_state_to_round}), with the encoded state being
a $|+\rangle$.  By contrast, the output of Steane's circuit-based
procedure is an encoded $|0\rangle$.  The difference between our
protocol and Steane's is due to the presence of the extra horizontal
bond between columns $7$ and $8$, which effects an encoded Hadamard
operation.  This will be compensated by a subsequent encoded Hadamard
operation, described below.

\subsubsection{Telecorrector creation}

To perform error correction we need to interact the data in our
cluster-state computer with the ancilla states in order to extract the
error syndrome.  We do this using a special cluster state which we
call a \emph{telecorrector}, which incorporates both multiple ancilla
states, as well as the cluster-based machinery to effect the necessary
interactions.  The telecorrector arises by clusterizing Steane's
protocol, but, as we shall describe, the cluster protocol enables
several modifications to improve the quality of the syndrome
extraction.  As in the previous section, our description is adapted to
the Steane $7$-qubit code, but is easily modified for many other
CSS codes.

Our clusterized method of syndrome extraction is based on the
following quantum circuit for syndrome extraction
\begin{equation} \label{eq:steane_extraction}
\epsfxsize=7cm \epsfbox{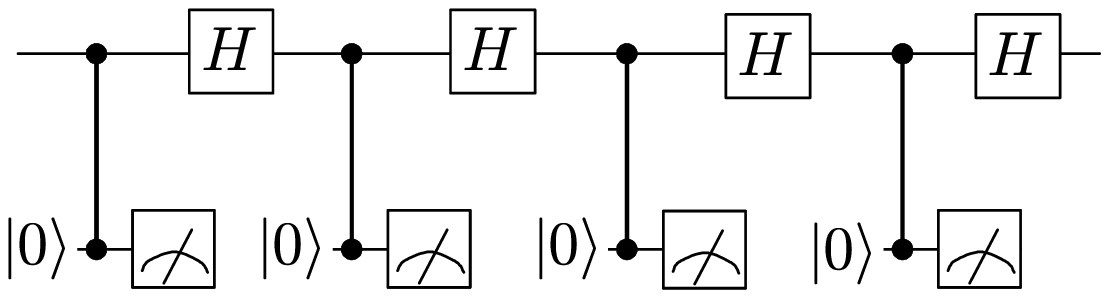}
\end{equation}
where operations are being performed on encoded qubits, $|0\rangle$ is
fault-tolerant ancilla creation, and the measurement is a transversal
$X$ basis measurement on constituent physical qubits in the code.
Circuit~(\ref{eq:steane_extraction}) is analogous to Steane's circuit,
except that the number of syndrome extractions is fixed, and the
syndrome extractions are performed in a different order: $X, Z, X, Z$,
in contrast to Steane, who extracts all $X$ information first,
followed by all $Z$. The reasons for these differences are explained
below.

Telecorrector creation begins with the creation of $7$ copies of
the following state:
\begin{equation} \label{eq:telemodule}
\epsfxsize=5.5cm \epsfbox{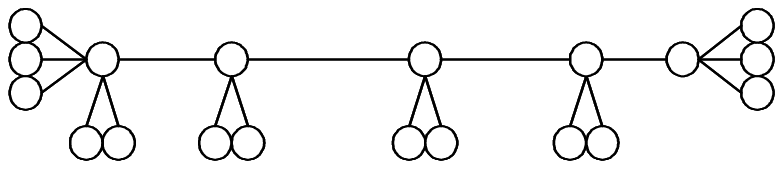}
\end{equation}
This state can be created in the obvious way using postselected
microcluster fusion.  The leaves on the left-hand end will
eventually be used to attach to a single qubit of the encoded data
using parallel fusion.  The leaves and root node on the right-hand
end will contain the output of this round of error-correction, and
become the input to the next round of error-correction.  The
remaining leaves will be used to fuse to ancilla states.

Simultaneous with the creation of Cluster (\ref{eq:telemodule}), we
create four verified ancilla states, using the technique described in
Section~\ref{sec:ancilla_creation}.  We then fuse the ancillas with
the leaves on Cluster (\ref{eq:telemodule}) to create the state:
\begin{equation} \label{eq:telecorrector}
\hspace{-1ex} \epsfxsize=9cm \epsfbox{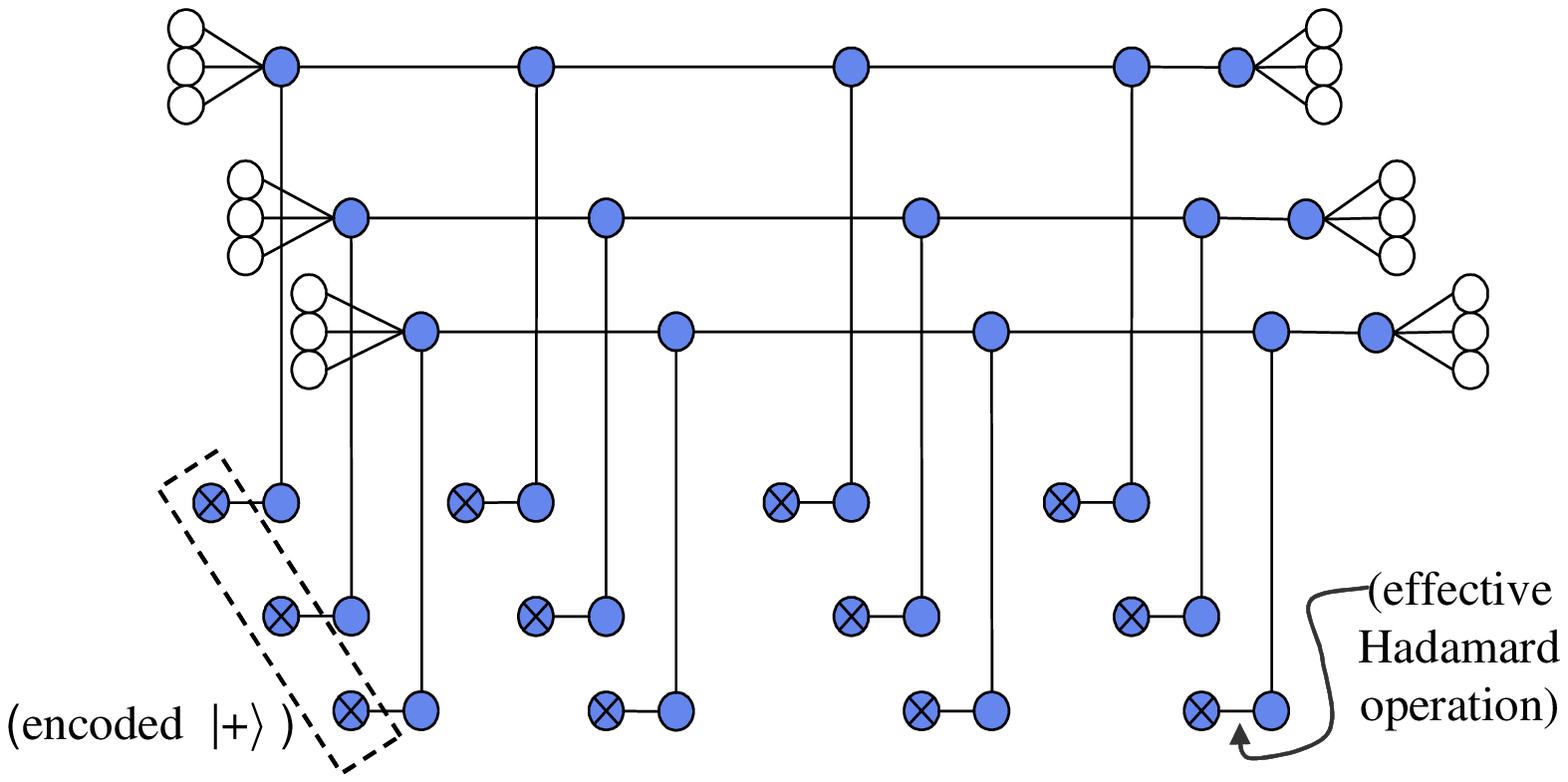}
\end{equation}
(The meaning of the shaded qubits will be explained below).  Note that
we have illustrated this as though only three qubits are involved in
the code: the case of $7$ (or more) qubit codes is similar, but the
diagram would be larger and more cluttered.

The next step is to measure all the shaded qubits in Cluster
(\ref{eq:telecorrector}) in the $X$ basis, leaving only the left-most
and right-most leaves, for later use in attaching the data, and future
rounds of error-correction.

Applying the propagation rules for the Pauli frame, it can be shown
that the pattern of measurement outcomes from the shaded qubits
completely determines whether or not the repeated syndrome
measurements will agree.

This is a remarkable property, since it enables us to determine
whether the repeated syndrome measurements will agree \emph{before}
the state has even interacted with the data.  Furthermore, we can take
advantage of this by postselecting on obtaining a set of measurement
outcomes that ensure this \emph{preagreeing syndrome} property.  We
call the postselected state with this preagreeing syndrome property
the \emph{telecorrector}.

Once prepared, we use parallel fusion to attach the telecorrector to
the data, and then $X$ basis measurements to complete this part of the
cluster-state computation.  Standard propagation rules are used to
update the Pauli frame, and to determine the final syndrome extracted
from this procedure.  We describe in the next section how this
syndrome information is decoded in order to perform correction.

The preagreeing syndrome property is responsible for the different
number and order of syndrome extractions in our protocol as compared
with Steane's.  Steane needs to extract many syndromes (more as the
code gets larger) in order to make it likely that some large subset of
those syndromes agree. In any round where syndromes don't agree,
correction cannot take place, and the round just adds more noise to
the data. We avoid this issue by using the preagreeing syndrome
property, thus reducing the number of locations at which noise can be
introduced into the data.

The preagreeing syndrome property also accounts for the order in which
we extract syndromes.  In Steane's protocol, the order of syndrome
extractions is all $X$ extractions in succession followed by all $Z$
extractions, so as to maximize the chance of obtaining syndromes that
agree. By extracting syndromes in the order $X, Z, X, Z$ we reduce the
chance of agreeing syndromes (for a small cost in resource usage) but
gain the ability to detect and postselect against additional types of
noise. In particular, $X$ errors that propagate from the second
 ancilla to become $X$ errors on the data
will be detectable via a disagreement of the first and third
syndromes. Likewise, $X$ errors that propagate from the third ancilla
to become $Z$ errors on the data will cause the second and fourth
syndromes to disagree.

\subsubsection{Reduction of fusion gate failure and photon loss to Pauli
  errors}

During the preparation of the ancilla and telecorrector states we used
postselection to avoid dealing with fusion gate failure and photon
loss.  This has the advantage both of improving the threshold, and
also means that our simulations don't need to describe these errors.
However, when the telecorrector is joined to the data, it is no longer
possible to postselect against these types of error, and we must find
some way of modeling them in our simulations.

By following a suitable experimental protocol, it turns out that both
these types of errors can be reduced to a (located) Pauli-type error,
which we already know how to model in our simulations.  The purpose of
this section is to describe this reduction.

In practice, we believe the protocol for reduction we describe is
likely to slightly worsen the behaviour of the cluster-state
computation.  The reason for introducing the reduction is therefore
not to improve the threshold, but rather to simplify our simulations,
and the statistics that we gather.  In actual experiments, the special
steps in the protocol described in this section would not need to be
performed, and the threshold would be slightly higher than our
simulations indicate.

We begin with a description of how we treat fusion gate failure.  The
discussion of photon loss will follow similar lines.

Suppose we are attempting to connect the telecorrector to the data
using parallel fusion, and all attempts fail, for a particular
horizontal row of qubits. The result of such a failure will be missing
horizontal bonds in the cluster. That is, instead of obtaining the
desired cluster
\begin{equation} \label{eq:join1}
\epsfxsize=3cm\epsfbox{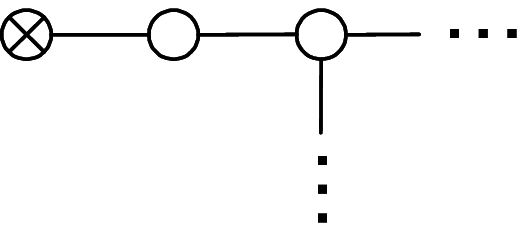}
\end{equation}
we obtain
\begin{equation} \label{eq:join2}
\epsfxsize=3cm\epsfbox{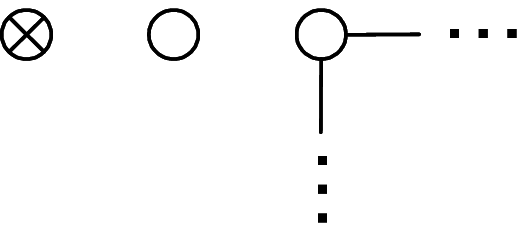},
\end{equation}
where the crossed node indicates a root node of the input data. We can
think of this as two located (i.e., known by the experimenter) {\sc
cphase} errors. (Note that in reality, the central bare node in
Equation~(\ref{eq:join2}) is not present, due to the destructive
measurement occurring with fusion failure. For the sake of the present
argument, we shall imagine that the experimenter has brought in a new
$|+\rangle$ state in this instance.) Unfortunately, our error model
for simulations doesn't allow us to describe {\sc cphase} errors
directly. Although we could imagine adding such an error to our list
of possible error types, the propagation rules turn out to be rather
complex, and we wish to avoid this if possible.

Suppose, however, that when parallel fusion fails, the experimenter
does the following:
\begin{itemize}
\item Depolarizes all three nodes of Equation~(\ref{eq:join2}) (that
is, replaces the state of the three qubits by the maximally mixed
state).

\item Notes the location of the row in which the failure occurred, for
use in
  decoding.

\item Carries out the rules for propagating the Pauli frame, as though
  the fusion gate had succeeded, and the horizontal bonds created.

\end{itemize}
Once the experimenter has performed the depolarization, it does not
make any physical difference whether the {\sc cphase} errors occurred
or not, and so we can imagine they have not occurred.  The only
remaining errors are Pauli-type errors, and so can be simulated in the
standard way.

Note that the effect of the intentional depolarization as described is
to randomize the results of later measurements performed on the three
qubits. Thus, our reduction of fusion failures to Pauli errors could
be equivalently achieved by the experimenter randomizing the
measurement results, without actually performing the depolarization.
Or simpler still, as a consequence of the propagation rules for the
Pauli frame, the randomization of the three measurement results could
be replaced by a randomization of the Pauli frame of the left-hand
root note. Thus, in our simulation, when a failed parallel fusion
between the data and telecorrector occurs, we simply randomize the
description of the Pauli frame error of the left-hand root node. This
completes the description of our procedure for modeling parallel
fusion failure when attaching the telecorrector to the data.

Consider now the case of photon loss.  Suppose a photon loss is
detected after fusion of the data and the telecorrector.  Recall from
Section~\ref{sec:physical_setting} that photon loss may occur in a
number of physically distinct ways: immediately after Bell pair
creation, before a memory step, before a fusion gate, and between the
fusion gate and measurement.  The effect of the photon loss may depend
on which of these possible ways it arose during the computation. In
particular, the effect may be described in the various cases as either
{\sc cphase} errors as in the case of failed fusion, or simply a
successfully created cluster followed by a single photon-loss error.
The experimenter would not know which of these cases had occurred.

To cope with this, we modify the protocol so that the effects (in any
of these cases) can be simulated by a Pauli-type error.  In the
modified protocol the experimenter does the following after detecting
a photon loss immediately after the data and telecorrector have been
fused.

\begin{itemize}
\item If a missing photon has been detected, the experimenter
  randomizes the local Pauli frame of the corresponding data qubit,
  i.e., the left-hand root node.

\item Notes the location at which the photon loss occurred, for use in
  decoding.

\item Carries out the rules for propagating the Pauli frame as though
  the horizontal bond between data and telecorrector had been
  successfully created.

\end{itemize}
This is simulated in the obvious way: when a photon loss is detected
after fusion of data and telecorrector, we randomize the Pauli frame
error of the corresponding data qubit, and apply the standard
propagation rules\footnote{One slight simplification we make in our
  simulations is to assume that photon loss may occur even following
  \emph{failed} fusion gates.  This can only worsen the thresholds
  obtained from our simulations, but the effect is negligible.}.  The
justification for following this procedure is very similar to fusion
gate failure, but requires the consideration of more separate cases,
corresponding to the different possible points of photon loss.  We
omit the details.

Note that a significant advantage of the frequent measurements
performed in the cluster model is that photon loss is detected before
it has a chance to propagate to adversely affect other parts of the
computation.  This is particularly useful as postselection can be used
to ensure that ancillas are free of photon loss noise, which helps
improve the threshold.

\subsubsection{Decoding}
\label{subsec:decoding}

We use a non-standard technique for syndrome decoding, designed to
take advantage of the knowledge the experimenter has of the locations
of errors caused by photon loss and nondeterminism. Our technique is a
maximum likelihood procedure for decoding arbitrary combinations of
located and unlocated errors.

We take advantage of the fact (see, e.g., Exercise~10.45 on
page~467 of~\cite{Nielsen00a}) that a code able to correct $t$
unlocated errors is also able to correct $2t$ located errors.
This is particularly advantageous for optical cluster-state
computation, since parallel fusion failure and photon loss errors
are likely the dominant types of noise.

Both the codes we will use in simulations (Steane 7-qubit and Golay
23-qubit) are CSS codes with the property that decoding of the $X$ and
$Z$ errors can be performed separately using an identical procedure.
Our description will be for the case of $X$ decoding; the $Z$ follows
similar lines.

The decoding routine has the following inputs: the measured $X$-error
syndrome, obtained by applying the classical parity check matrix to
the vector of total errors of the ancilla measurement outcomes; and a
list of locations (qubit indices within the code block) at which
located errors have occurred during the round. The outputs to the
decoding routine are: a list of locations where $X$ flips should be
made in order to correct the data; and a flag signaling a
\emph{located crash}.

The located crash flag indicates that the correction has likely
failed, and the logical state of the data has effectively experienced
a random $X$ operation (i.e. an $X$ \emph{crash}). This situation
arises when two different patterns of $X$ errors are found to have
equal maximum likelihood, but differ from each other by a logical $X$
operation. The located crash flag is not used directly, but will be
used to assist decoding at the next level of concatenation, by
identifying encoded blocks which are known to have experienced an
error. By feeding information in this way to higher levels of
concatenation, we are increasing the overall noise-threshold
performance of the protocol.

Before describing our maximum likelihood decoder, we first give a
simple model for the relative likelihood of errors. The total $X$
error pattern on the data is a product of $X$s due to unlocated
errors, and $X$s due to located errors. The measured syndrome is
assumed to be the bitwise exclusive or of the syndromes of the two
error patterns. The likelihood of a pattern of unlocated $X$s is
assumed to be a decreasing function of weight, but not a function of
how the errors are positioned. The likelihood of a pattern of $X$
errors due to located errors is uniform across all patterns which have
$I$ wherever located errors have not occurred. This is due to our
reduction of located errors to depolarization. For example, if located
errors have occurred on three qubits, then the resulting $X$ error
pattern on those qubits due to the located errors will be either
$III$, $IIX$, \dots, $XXX$ with equal probability, and $I$ on other
qubits.

To decode, we loop over all possible values for the located error
pattern, and for each one we determine the most likely unlocated error
pattern. For a particular located error pattern, the most likely
unlocated error pattern is found by first finding its syndrome, by
taking the exclusive or of the measured syndrome with the syndrome of
the located error pattern. Then from this syndrome, the most likely
unlocated error pattern is found via a standard decoding array. As the
loop is repeated over all located error patterns, we keep track of
which ``most likely unlocated error pattern'' has the overall minimum
weight, and is thus most likely overall. If this minimum is unique,
then the data is corrected\footnote{Note that in both the
  cluster-based and deterministic protocols we don't ever physically
  apply the corrections.  Instead, by ``correcting'' the data we
  simply mean that we keep track of the corrections that must be
  applied, and propagate them forward through the computation to be
  compensated at the end, much as we treat the Pauli frame.
}  by first
correcting for this minimum weight unlocated error pattern, then
correcting for the corresponding located error pattern. The located
crash flag is set to ``false''.

Otherwise, if the minimum is not unique, we arbitrarily choose one of
the minimum weight patterns and corresponding located error pattern,
and correct accordingly. We compare the correction performed against
the corrections associated with each of the other minima. If they are
all equivalent up to stabilizer operations of the code, then we set
the located crash flag to ``false''. Otherwise we set the located
crash flag to ``true''.

\subsubsection{Results of the optical cluster simulation}
\label{subsubsec:ocluster_results}

To determine the threshold for a concatenated error correction
protocol, we must analyse how the effective error rates vary as more
levels of concatenation are added. We now give results of this
analysis for the lowest level of concatenation -- the cluster-based
protocol of Subsection~\ref{subsec:ft_protocol_cluster}. We simulated
this protocol with the aim of categorizing the function that maps the
physical noise parameters $(\epsilon, \gamma)$ to the logical error
rates, or {\em crash} rates, defined below. Likewise, in
subsection~\ref{subsec:ft_protocol_deterministic} we describe
simulations which categorize the similar function for a deterministic
circuit-based protocol, representing higher levels of concatenation.
In subsection~\ref{subsec:results}, the results are combined to give
the threshold region for concatenated cluster-state optical quantum
computing.

Two of the authors, CMD and HLH, each created a version of the
simulator, and no program code was shared between the two versions.
This duplication was done so that agreement between the results of the
two simulators could act as a verification that the simulators were
bug free. The programming languages C++ and C were used for the most
part (and to a lesser extent, Python and {\sc Matlab}).

At the end of a round of simulated cluster-based error correction, we
say that the round has caused a {\em located crash} whenever either
the $X$ or $Z$ decoding steps in that round has reported a located
crash. Note that the imagined experimenter would be aware of located
crashes occurring. In addition, we define an {\em unlocated crash} as
follows. We take the pattern of total Pauli errors on the root nodes
of the data, and consider the result of a perfect (noise-free) round
of correction. If perfect correction would result in a pattern of
Pauli errors equivalent to an encoded $X$, $Y$, or $Z$ Pauli
operation, then we say the data has experienced an unlocated crash.
Note that errors on the leaves of the data are not taken into account
when we test for an unlocated crash. Such errors are not completely
ignored, as they will instead propagate to the next round of error
correction and affect the next crash rate test.

We performed four separate sets of simulations, in order to compare
the use of two different codes and two different settings for memory
noise. The four configurations were: Steane 7-qubit code with and
without memory noise enabled; and the Golay 23-qubit code with and
without memory noise enabled. In the cases where memory noise was
disabled, we did not apply photon loss or depolarization noise during
memory operations. Comparing the results with memory noise enabled and
disabled gives an indication of how significant the effect of memory
noise is on the performance of the protocol.

For each of the four configurations noted above, we chose a number of
settings for the noise parameters $(\epsilon,\gamma)$, and for each we
ran a many-trial Monte Carlo simulation. Each trial of the Monte Carlo
simulation consisted of two successive rounds of the error correction
protocol, and the outcome of the trial was determined by whether the
second of the two rounds caused a crash. The purpose of the first
``warm-up'' round is to reduce the transient effects due to our choice
of (noise free) initial conditions. We found that including more than
one warm-up round did not make a statistically significant change to
the results. However, not including a warm-up round did affect results
considerably.

The number of parallel attempts per leaf-to-leaf fusion during ancilla
cluster creation and during the joining of ancilla to the
telecorrector was set to $3$ throughout. Recall, as far as the noise
performance is concerned, the fewer attempts per leaf-to-leaf fusion
the better during the above mentioned cluster building steps. However
we chose $3$ as a compromise between noise performance and resource
usage.

The number of parallel attempts per leaf-to-leaf fusion when joining a
telecorrector to the data cluster was set at five throughout. Here,
fewer attempts is not necessarily better for noise performance,
because when all attempts fail, a located error is introduced to the
data. We found that using any figure above five gave a consistently
worse final threshold, whereas a figure less than five gave a worse
threshold for small values of $\gamma$ but a slightly better threshold
for larger $\gamma$ values.

The various outcomes of each trial are tallied as follows. For all the
trials for which the first round does not cause a crash, we count: (1)
the number $N_U$ of trials for which the second round causes an
unlocated crash but not a located crash, (2) the number $N_L$ of
trials for which the second round causes a located crash, and (3) the
number $N_N$ of trials for which no crashes occur.

>From the values $N_U$, $N_L$, and $N_N$, the unlocated and located
crash rates $E$ and $\Gamma$ are estimated as follows:
\begin{eqnarray}
E&=&\frac{N_U}{N_U+N_N}, \\
\Gamma&=&\frac{N_L}{N_U+N_N+N_L}.
\end{eqnarray}
Note that we omit $N_L$ from the denominator of $E$ since we only
compute the unlocated crash rate conditional on no located crash
having occurred.  The estimated standard error for $E$ and $\Gamma$
respectively are
\begin{eqnarray}
\sigma_E&=&\frac{\sqrt{N_U}}{N_U+N_N}, \\
\sigma_\Gamma&=&\frac{\sqrt{N_L}}{N_U+N_N+N_L}.
\end{eqnarray}
Both these expressions arise from the fact that if we sample $N$ times
to estimate the probability $p$ of an event occurring, then the
standard deviation in the estimate is $\sqrt{p(1-p)/N}$.  When $p$ is
small, as it is in our case, we may neglect the $1-p$ term to obtain a
standard deviation of $\sqrt{p/N}$.

The two versions of the simulator program code were compared as
follows. For $65$ different settings of $(\epsilon,\gamma)$, the
values $(E,\Gamma)$ were estimated from each simulator using a sample
size of at least $10^6$. This was done for the $7$-qubit code, both
for memory noise disabled and enabled. For the resulting 130 different
values, we compared the results obtained by the two simulators, and
the largest difference observed was $3.1$ times the estimated standard
error. In other words, the two independently-created simulators showed
excellent agreement, and this provides additional evidence that they
are free of serious bugs.

One of the versions of the simulator code was used to gather final
results. We denote the particular choices of the input noise
parameters as $(\epsilon_i,\gamma_i)$, $i=1,\dots,D$, the
corresponding crash rate estimates as $E_i$ and $\Gamma_i$, and the
corresponding standard errors as $\sigma^E_i$ and $\sigma^\Gamma_i$.
For the $7$-qubit code, approximately $10^7$ samples were run for each
of 59 different settings of the noise parameters
$(\epsilon_i,\gamma_i)$, for both disabled and enabled memory noise.
(Note that the particular choices used for $(\epsilon_i,\gamma_i)$ are
shown as small circles on the threshold plots in the final results
subsection, \ref{subsec:results}).

For the $23$-qubit code, samples were gathered for 43 different noise
parameter settings, for both enabled and disabled memory noise. The
sample sizes ranged from $4\times 10^4$ to $3\times 10^7$ for disabled
memory noise, and from $3\times 10^5$ to $2 \times 10^7$ for enabled
memory noise. The smaller sample sizes correspond to highest noise
rate settings, where the simulation becomes much slower (due to noisy
ancilla and telecorrectors being discarded more often, an effect which
is much more pronounced for the $23$-qubit code compared with the
$7$-qubit code).

We fit polynomials to the data using weighted least-squares fitting. A
polynomial $E(\epsilon,\gamma)$ is fitted to the values $E_i$ by
minimizing the following residual:
\begin{equation}
R_E = \sum_{i=1}^D \frac{\left(E(\epsilon_i,\gamma_i) - E_i\right)^2}{
(\sigma^E_i)^2}.
\end{equation}
Likewise, the polynomial $\Gamma(\epsilon,\gamma)$ is fitted to the
values $\Gamma_i$ by minimizing the residual:
\begin{equation}
R_\Gamma = \sum_{i=1}^D \frac{\left(\Gamma(\epsilon_i,\gamma_i) -
\Gamma_i\right)^2}{ (\sigma^\Gamma_i)^2}.
\end{equation}

All terms up to order five were included in the polynomial
$E(\epsilon,\gamma)$, with the exception of terms of order 0 in
$\epsilon$.  The reason for the excluded terms is that we know
$E(0,\gamma)=0$. In the polynomial $\Gamma(\epsilon,\gamma)$, all
terms up to order six were included for the 23-qubit code results, and
terms up to order five were included for the 7-qubit code. In each
case the chosen orders of five or six were the minimum that gave a
``good'' fit to the data for all four configurations of code and
memory noise. We considered a good fit to be when the residual divided
by the number of data points $D$ was roughly of order $1$ (in practice
the value ranged from $0.42$ to $1.43$ for the eight polynomials
fitted). Such a condition indicates that the differences between the
observed values and the fitted polynomial could reasonably be
accounted for solely by the errors due to the finite sample size.

It would be rather cumbersome to give all the fitted polynomials
obtained, in isolation from the procedure in
Subsection~\ref{subsec:results} to convert this information to a
threshold region. Rather, as an example of the results, we give the
coefficients of the polynomial $E(\epsilon,\gamma)$ for the case of
the 23-qubit Golay code with memory noise enabled, in
Table~\ref{table:poly}.

\begin{table}
\begin{tabular}{r|l}
Monomial &Coefficient \\ \hline
 $\epsilon  $&     $0.003357$ \\
 $\epsilon \gamma $&        $2209$ \\
 $\epsilon \gamma^2 $&  $-3.630\times10^6$ \\
 $\epsilon \gamma^3 $&$   1.868\times10^9$ \\
 $\epsilon \gamma^4 $&$  -8.421\times10^{10}$ \\
 $\epsilon^2  $&$  2009  $\\
 $\epsilon^2 \gamma $&$  -2.133\times10^7$ \\
 $\epsilon^2 \gamma^2 $&$   2.979\times10^{10} $\\
 $\epsilon^2 \gamma^3 $&$   -2.573\times10^{12} $\\
 $\epsilon^3  $& $  -3.578\times10^7$ \\
 $ \epsilon^3 \gamma$&$  2.348\times10^{11} $\\
 $ \epsilon^3 \gamma^2 $&$ -2.9574\times10^{13} $\\
 $ \epsilon^4 $&$ 7.098\times10^{11} $\\
 $\epsilon^4 \gamma $&$   -2.341\times10^{14}$ \\
 $ \epsilon^5  $&$  -2.472\times10^{14}$
  \end{tabular}
\caption{The polynomial $E(\epsilon,\gamma)$ as fitted to the
unlocated crash rate data, for the cluster-state protocol, using the
23-qubit code with memory noise enabled.}\label{table:poly}
\end{table}

\subsection{The deterministic protocol}
\label{subsec:ft_protocol_deterministic}

In this subsection we describe the simulation of our deterministic
(circuit based) error-correction protocol. Much of the detail of the
protocol is given in Appendix~\ref{sec:telecorrection}. The main
purpose of the present subsection is to explain how the deterministic
protocol fits together with the cluster-based protocol, describe the
effective noise model used for simulating the deterministic protocol,
and to give the methods and results of these simulations.

\subsubsection{Concatenation of protocols}
To perform a threshold analysis, one usually imagines that a
fault-tolerant error-correction protocol is concatenated with itself
many times. That is, the encoded qubits corrected by the circuit at
the lowest level of concatenation are themselves used to build up a
circuit for error correction at a higher level of encoding, and so on.
Then, by definition, a physical error rate is ``below the threshold''
if the rate of logical errors (crashes) at the highest level of
encoding can be reduced arbitrarily close to zero by using
sufficiently many levels of concatenation. Usually, to simplify
analysis, the error correction circuit and noise model at every level
are taken to be identical, and the rate of noise per gate at one level
is taken to be the rate of crashes per error-correction round at the
next lowest level. With these set of assumptions, the task of
determining if a particular noise rate is below the threshold becomes
that of simulating just the lowest level of concatenation, and testing
whether the crash rate is below the physical noise rate.

In the quantum computation that we are simulating, only the lowest
level of concatenation uses the cluster based protocol described in
subsection~\ref{subsec:ft_protocol_cluster}. For the second and higher
levels of concatenation, we can effectively regard it as though a
circuit-based deterministic protocol is being used, since the encoded
gates available to higher levels of concatenation are deterministic.
Steane's fault-tolerant protocol would be a suitable choice for the
higher levels of concatenation, but rather we have chosen to use the
telecorrection protocol of Appendix~\ref{sec:telecorrection} for the
reasons we outline in that appendix.

To motivate the ensuing description of the effective noise model used
in the simulations of the deterministic protocol, we discuss the way
in which a gate or other operation at one level of concatenation is
built from the error-correction protocol at the next lower level of
concatenation. The operations used in the telecorrector circuit are:
{\sc cphase} and {\sc cnot} gates; preparation of $|0\rangle$ and
$|+\rangle$; $X$-basis measurements; and memory. First we discuss how
these operations in the level $L\ge 3$ of concatenation are built from
level $L-1$.

The memory operation at level $L$ is simply one round of error
correction at level $L-1$. Accordingly, in our noise model for memory
operations at level $L$ , the various noise types are introduced with
probabilities given by the crash rates of a round of level $L-1$ error
correction (the details will be made clear later).

Each of the two types of gates used in the telecorrection circuit at
level $L$ are implemented by first applying a round of error
correction to the inputs of the gate, then applying the encoded gate
consisting of the level $(L-1)$ gate applied transversally to each
qubit in the code. The error correction stage contains many more gates
than the actual encoded gate, thus we assume that the majority of the
noise introduced by a gate is due to the error correction step.
Accordingly, in our noise model for gates at level $L$, noise is
introduced to each of the gate inputs according to the model for
memory noise (that is, again given by the crash rates of a level $L-1$
correction round).

The preparation of $|0\rangle$ or $|+\rangle$ at level $L$ can be
implemented by preparing the level $L-1$ state transversally on each
qubit in the code, followed by a round of error correction. Again, we
assume that most of the noise is due to the error-correction step, and
at level $L$ our model introduces noise after preparation operations
according to the model for a step of memory noise.

An $X$-basis measurement at level $L$ is implemented by measuring each
level $L-1$ qubit transversally in the $X$ basis, then performing
classical error correction on the results. So in contrast to the other
operations, measurement does not involve a quantum error correction
round at the lower level, but rather a noise free classical
correction. Accordingly, our noise model assigns a much lower rate of
noise to measurements at level $L$ relative to the rates of the other
operations.

Similar arguments can be made for operations at level 2 built from the
cluster protocol of the lowest level. Thus, we will take the rates of
noise introduced to gates, memory and preparation at level 2 equal to
the crash rates due to a round of clusterized error correction, but
make the noise due to measurement significantly less.

We now specifically state the effective noise model used at a level
$L\ge2$ of concatenation, following the arguments above.

\subsubsection{Effective noise model}

In the simulation of the level $L\ge 2$ error correction circuit, we
model the encoded qubits that this circuit acts upon as though they
were physical qubits. That is, at every stage of the simulation of the
circuit, the error description is a Pauli error, $I$, $X$, $Y$ or $Z$,
associated with each of the qubits. The details of the errors on lower
level qubits are not directly simulated. As in the cluster-based
protocol, the circuit is divided into time-steps. Each qubit in the
level-$L$ circuit can undergo one operation per time-step. The length
of a time-step corresponds to the time taken for a complete round of
error correction and an encoded operation to be performed at level
$L-1$.

Our model involves four types of noise, unlocated $X$ and $Z$ Pauli
errors, and located $X$ and $Z$ Pauli errors. Unlocated and located
errors are designed to represent the unlocated and located crashes
occurring at level $L-1$. When a qubit experiences an unlocated $X$
Pauli error, it undergoes an $X$ operation, unknown to the
experimenter, and similarly for unlocated $Z$ Pauli errors. When a
qubit experiences a located $X$ error, it undergoes an $X$ operation
with probability $1/2$. The experimenter will know that a located $X$
error has occurred, but not whether the corresponding $X$ Pauli error
has actually been applied. $Z$ located errors are similar.

When we say that unlocated noise is applied with a probability $p$, we
mean that both unlocated $X$ and $Z$ Pauli errors are applied with
equal probability, and independently, such that the total probability
that an error was applied is $p$. Similarly for located noise applied
with probability $q$. We choose this model of independent $X$ and $Z$
errors because of a numerical observation that the rate of $Y$ crashes
is much less than the combined rate of $X$ and $Z$ crashes, for both
our cluster-based and deterministic protocols. Although observed $X$
and $Z$ crash rates are not entirely independent, we have nonetheless
chosen an independent noise model, which empirically appears to
provide a good approximation to the observed behaviour.

We now describe how noise is introduced by each operation. Let $p$ and
$q$ be the rates of unlocated and located crashes respectively for an
error correction round at level $L-1$.
\begin{itemize}
\item Memory and gates: Before the gate or memory,
 the following noise is applied to the input qubit, or in the case of
 two-qubit gates is applied independently to each input.
 Unlocated noise is applied with probability $p$, and,
independently, located noise is applied with probability $q$.
\item Preparation: After the preparation, unlocated noise is applied
with probability $p$, and, independently, located noise is applied
with probability $q$.
\item Measurement: Before the measurement, unlocated noise is applied
with probability $p/10$, and, independently, located noise is applied
with probability $q/10$.
\end{itemize}

The value of one tenth for the relative strength of measurement noise
is somewhat arbitrary. In reality, the relative strength of
measurement noise compared with other noise types would decrease for
higher levels of concatenation. This is because, for higher levels of
concatenation, the implementation of an encoded gate (or other
non-measurement operation) becomes increasingly more complicated
compared to that of an encoded measurement. We obtained numerical
evidence to suggest that even after just one concatenation of the
deterministic protocol, the relative rate of crashes from an encoded
measurement was {\em less} than one tenth that of other operation
types. So, our choice to fix the value at $1/10$ is likely to be a
little pessimistic, but is nonetheless much more realistic than
setting equal noise strengths for all operation types.


\subsubsection{Telecorrection protocol}

We simulate the protocol described in
Appendix~\ref{sec:telecorrection}, in particular using the layout of
Circuit~(\ref{eq:telecorrector2}). We now briefly describe some
further pertinent details of the protocol not given in the appendix,
namely the circuit used for ancilla creation/verification, the
procedure for post-selection during telecorrection creation, and the
decoding procedure.

The circuit used to create and verify encoded $|0\rangle$ ancilla
states, denoted by the operation ``$|0\rangle$'' in
Circuit~(\ref{eq:telecorrector2}), uses the design of Steane. For
example, for the seven qubit code, the circuit is:
\begin{equation}
\epsfxsize=8.2cm \epsfbox{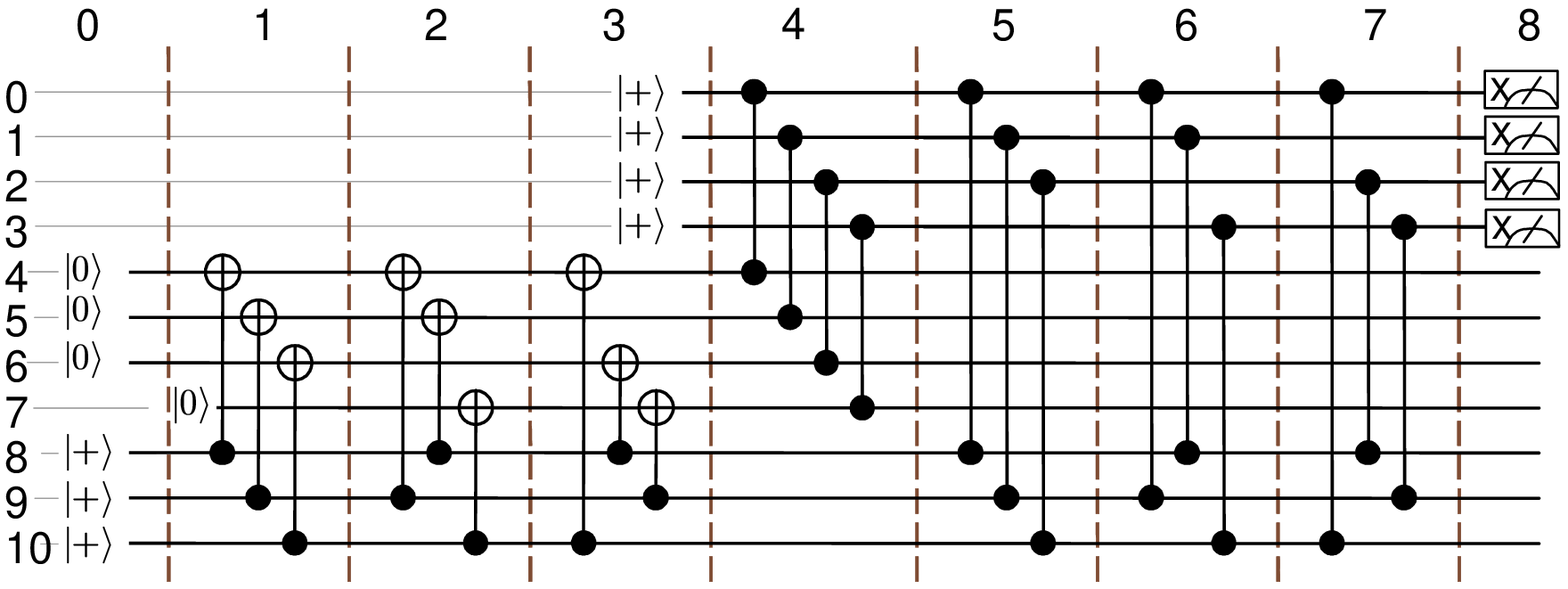}
\end{equation}
where the measurements are post-selected to have outcome ``0''.

Note that in the case of the 23-qubit code (not shown), our circuit
for ancilla creation and verification has the advantage of taking 8
fewer time steps than that used by Steane \cite{Steane03a} for the
same code. This is due to the fact that we start with a version of the
classical Golay code having a reordering of the 23 bits in the code.
By reordering bits in the code (i.e., permuting columns in the check
matrix) and then reexpressing the check matrix in standard form, it is
possible to change the maximum column and row weight of the check
matrix, which has the effect of changing the number of time steps in
the creation and verification circuits. After trying many random
bit-reorderings, we found that the number of time steps in the circuit
could be made as low as 17, compared with Steane's 25.

The telecorrector-creation part of the protocol, indicated by the
boxed region in Circuit~(\ref{eq:telecorrector2}), is performed many
times in parallel, and post-selected to give a successfully created
telecorrector state. Here, ``successful'' means that syndromes of like
type agree, and that no located noise occurred during the creation
circuit.

During the protocol, the data and one half of the telecorrector are
measured, in order to effectively apply two successive encoded
transport circuits. Each of the two encoded measurements consists of
$X$-basis measurements on each of the qubits in the code, followed by
classical error correction performed on the measurement results. In
each case, the correction procedure involves: (1) calculating the
syndrome associated with the measurement results, (2) determining
which of the individual measurement results within the encoded
measurement were subject to located noise, and (3) using the results
of the first two steps as input to the decoding procedure of
Subsection~\ref{subsec:decoding}.

\subsubsection{How we simulate the protocol}

A simulation trial begins with the state of the quantum computer being
noise-free. Thus, the description of the initial state is a Pauli
error of $I$ on each data qubit\footnote{Note that the Pauli frame,
used in the optical cluster protocol, does not form part of the
deterministic protocol. Thus we do not keep track of Pauli frame
errors when simulating the deterministic protocol.}. Then, some number
of repeated telecorrection rounds are simulated. As each operation in
the circuit is simulated, the Pauli error description of the qubits
are updated stochastically based on the unlocated noise model, and
Pauli errors are propagated as appropriate for the operation. The
propagation rules for each operation are:
\begin{itemize}%
\item Preparation: Pauli error is reset to $I$. %
\item Measurement: Measurement in the $X$ basis causes the $Z$ part of
the Pauli error on a qubit to propagate to the measurement result, and
the $X$ part of the error to be eliminated. %
\item {\sc cnot} gate: A Pauli error of $X^{x_t}Z^{z_t}$ on the target
and $X^{x_c}Z^{z_c}$ on the control are transformed as
\begin{eqnarray}
x_t' &=& x_t + x_c \\
z_t' &=& z_t \\
x_c' &=& x_c \\
z_x' &=& z_c + z_t.
\end{eqnarray}
\item {\sc cphase} gate: A Pauli error of $X^{x_1}Z^{z_1}$ and
$X^{x_2}Z^{z_2}$ on the two inputs are transformed as
\begin{eqnarray}
x_1' &=& x_1  \\
z_1' &=& z_1 + x_2 \\
x_2' &=& x_2 \\
z_2' &=& z_2 + x_1.
\end{eqnarray}
\end{itemize}%

To speed up simulations, located noise is not introduced where it will
later be post-selected away. Located noise which cannot be
post-selected away occurs due to the following operations in the
protocol: the transversal {\sc cphase} between the data and the one
half of the telecorrector; the memory step on the other half of the
telecorrector during the aforementioned transversal {\sc cphase}; and
the measurements of the data and one half of the telecorrector.  A
straightforward analysis of error locations shows that the effect of
all these located noise events is statistically equivalent to applying
a located error at the start of the round with a suitable probability.
We omit the details of this analysis, but note that for simplicity in
simulation we used this simplified error model.

\subsubsection{Results of simulating the deterministic protocol}
\label{subsubsec:det_results}

As in the simulations of the cluster-based protocol, we aim to
categorize the function which maps input noise parameters, in this
case the unlocated noise rate $p$ and located noise rate $q$, to the
logical error rates, being the unlocated crash rate $P$ and located
crash rate $Q$. From knowledge of this map for both the cluster-based
and deterministic protocol, the overall threshold region can be
determined.

Again we performed separate sets of simulations using the 7-qubit
Steane code with and without memory noise enabled, and using the
23-qubit Golay code with and without memory noise enabled. Note that
in the case where memory noise is disabled, we still apply memory
noise on the bottom half of the telecorrector during the timestep in
which the data and top half of the telecorrector are interacting with
the {\sc cphase} gate. This location in the circuit is where any
encoded gate would be performed between correction rounds, and so we
apply noise here in every circumstance so that the noise due to this
encoded operation is taken into account.

For a particular choice of code and memory noise setting, we chose a
number of settings for the parameters $(p,q)$, and for each we ran a
many-trial Monte Carlo simulation. As for the cluster-state
simulations, each trial of the Monte Carlo simulation consisted of two
successive rounds of the error correction protocol, with statistics
gathered on the rate of crashes introduced by the second round. Again,
including more than two rounds did not appear to affect results.

The definition of unlocated and located crashes for a round of the
deterministic protocol is virtually identical to that given in
subsection~\ref{subsubsec:ocluster_results}. Similarly, the tallies
$N_L$, $N_U$ and $N_N$ for the various trial outcomes share the same
definition as in subsection~\ref{subsubsec:ocluster_results}.

The unlocated and located crash rates $P$ and $Q$ are estimated as
follows:
\begin{eqnarray}
P &=& \frac{N_U}{N_U+N_N}, \\
Q &=& \frac{N_L}{N_U+N_N+N_L}.
\end{eqnarray}
The estimated standard error for each quantity is
\begin{eqnarray}
\sigma^P = \frac{\sqrt{N_U}}{N_U+N_N},\\
\sigma^Q = \frac{\sqrt{N_L}}{N_U+N_N+N_L}.
\end{eqnarray}

The results of two independently-written simulators were compared, as
in the case of the optical cluster state protocol, as a check on
whether the results were bug free. Estimates of the quantities $P$ and
$Q$ were compared between the two versions of the simulator, using the
$7$-qubit code and a sample size of approximately $10^6$, for 68
different noise settings with memory noise disabled and 86 different
noise settings for memory noise enabled. Comparisons of a lesser
sample size were also carried out for the $23$-qubit code. The largest
discrepancy found during all comparisons equated to $3.2$ times the
estimated standard deviation. Thus the two simulators showed excellent
agreement, and this provides additional evidence that they are free of
serious bugs.

Final results were gathered using one of the versions of the
simulator. Denote the choices of the input noise parameters as
$(p_i,q_i)$, $i=1,\dots,D$, the corresponding crash rate estimates as
$P_i$ and $Q_i$, and the corresponding standard errors as $\sigma^P_i$
and $\sigma^Q_i$. Polynomials were fitted to the data using weighted
least-squares fitting. A polynomial $P(p,q)$ is fitted to the values
$P_i$ by minimizing the following residual:
\begin{equation}
R_P = \sum_{i=1}^D \frac{\left(P(p_i,q_i) - P_i\right)^2}{
(\sigma^P_i)^2}.
\end{equation}
Likewise, the polynomial $Q(p,q)$ is fitted to the values $Q_i$ by
minimizing the residual:
\begin{equation}
R_Q = \sum_{i=1}^D \frac{\left(Q(p_i,q_i) - Q_i\right)^2}{
(\sigma^Q_i)^2}.
\end{equation}

All terms up to order six were included in the polynomial $P(p,q)$,
with the exception of terms of order 0 in $p$. In $Q(p,q)$, all terms
up to order five and eight respectively were included when using the
$7$ and $23$-qubit codes, except terms of order $0$ in $q$. The reason
for the excluded terms is that we know $P(0,q)=0$ and $Q(p,0)=0$. The
orders were chosen using a similar criteria as for optical cluster
protocol.

To present the results of the deterministic simulations, we calculate
a threshold region with respect to the noise parameters at the second
level of concatenation. (Thus, we are temporarily ignoring the effect
of the optical cluster protocol at the lowest level).
Define the map $g: (p,q)\rightarrow (P(p,q),Q(p,q))$, where $P$ and
$Q$ are the fitted polynomials. If $(p,q)$ are the effective unlocated
and located noise rates at the second level of concatenation, then the
unlocated and located crash rates at the $k$-th level may be estimated
by computing $g^{(k-1)}(p,q)$. Provided this tends towards $(0,0)$ as
$k\rightarrow \infty$ the point $(p,q)$ is inside the threshold region
for the deterministic protocol. It is possible to test many thousands
of points very quickly using this method, giving the threshold to high
resolution.

\begin{figure}
\begin{center}
\epsfxsize=8.7cm \epsfbox{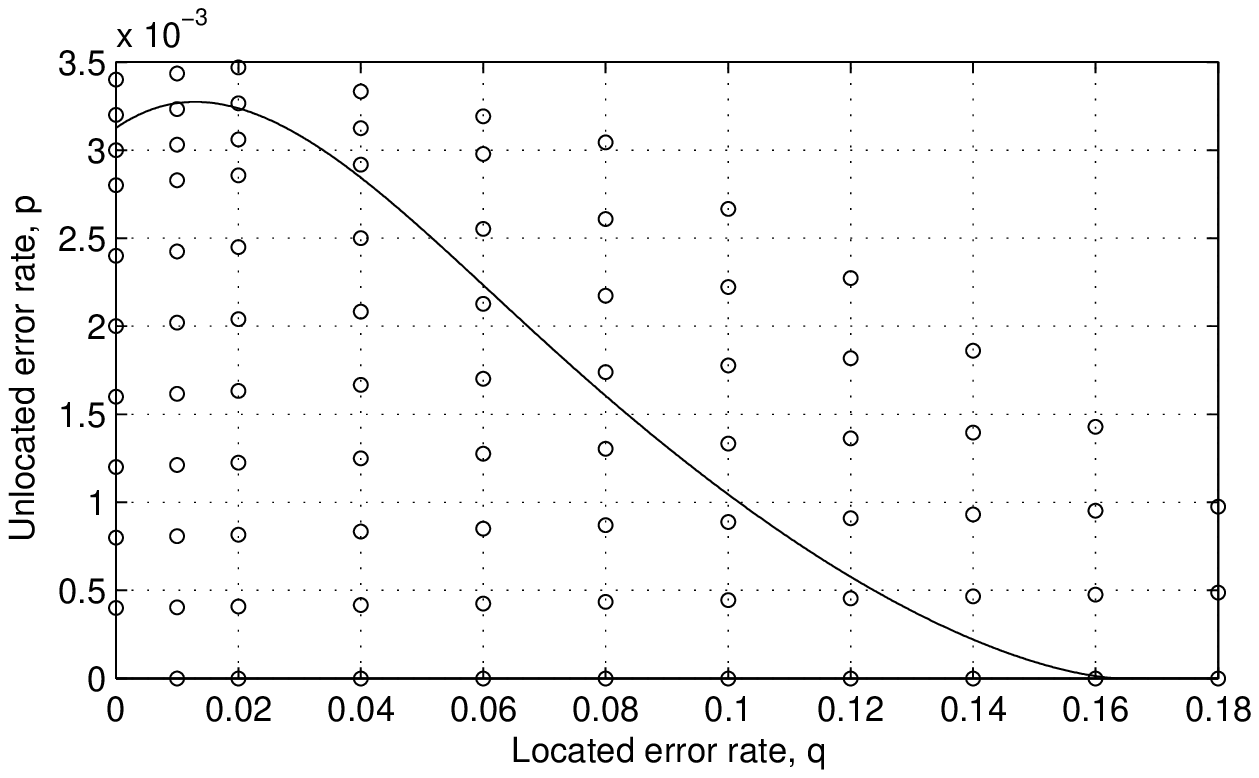}
\end{center}
\begin{center}
\epsfxsize=8.7cm \epsfbox{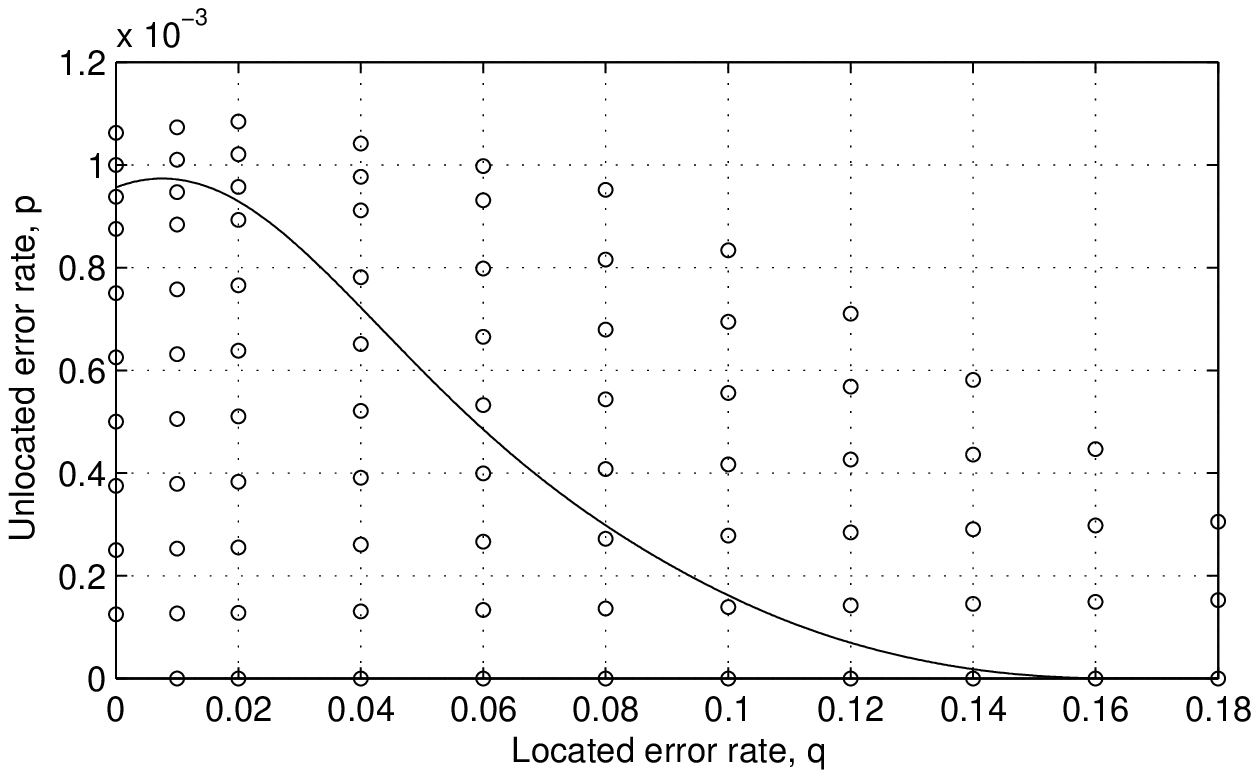}
\end{center}
\caption{Threshold region (below the solid line) for the deterministic protocol
  using the 7-qubit Steane code. Memory noise is disabled top, and
  enabled bottom. Circles indicate the noise parameter values for
  which the simulation was run.} \label{fig:det7}
\end{figure}

The threshold regions for the simulations using the 7 qubit code are
shown in Figure~\ref{fig:det7}. For each of the points $(p_i, q_i)$
shown by the circles, between $10^7$ and $2\times 10^7$ trials were
run.
\begin{figure}
\begin{center}
\epsfxsize=8.7cm \epsfbox{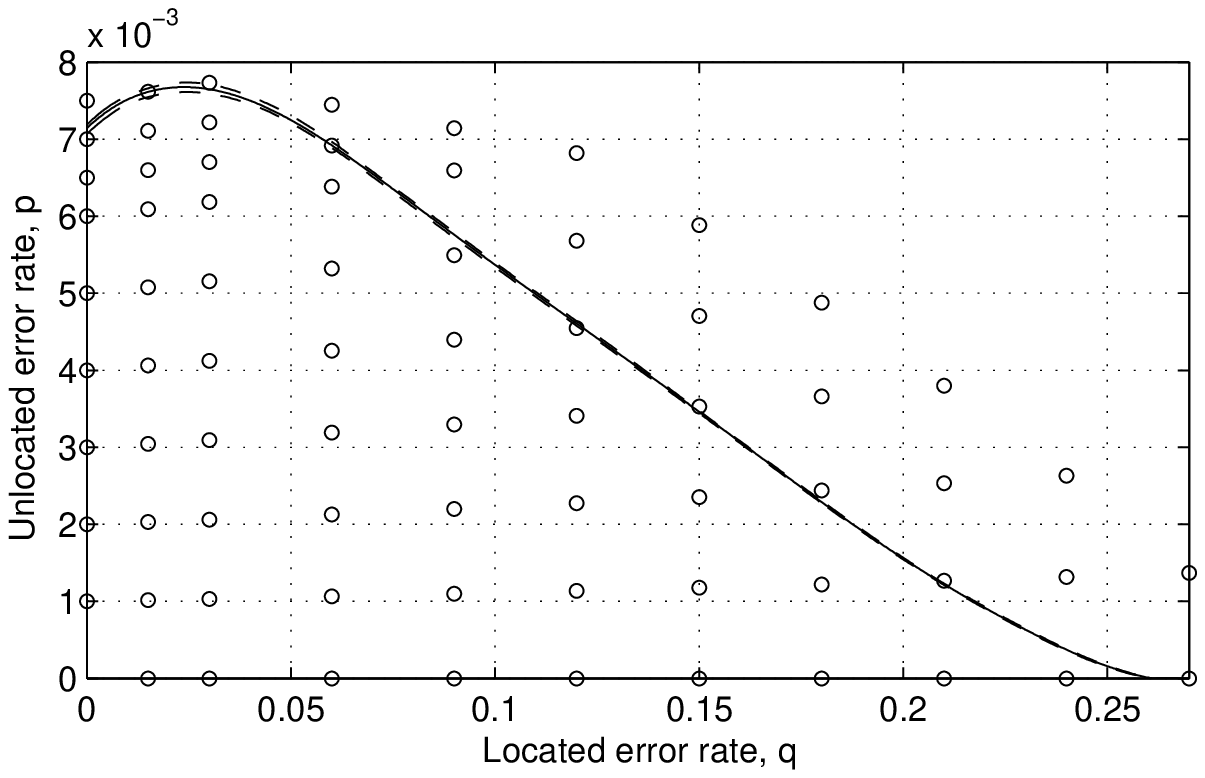}
\end{center}
\begin{center}
\epsfxsize=8.7cm \epsfbox{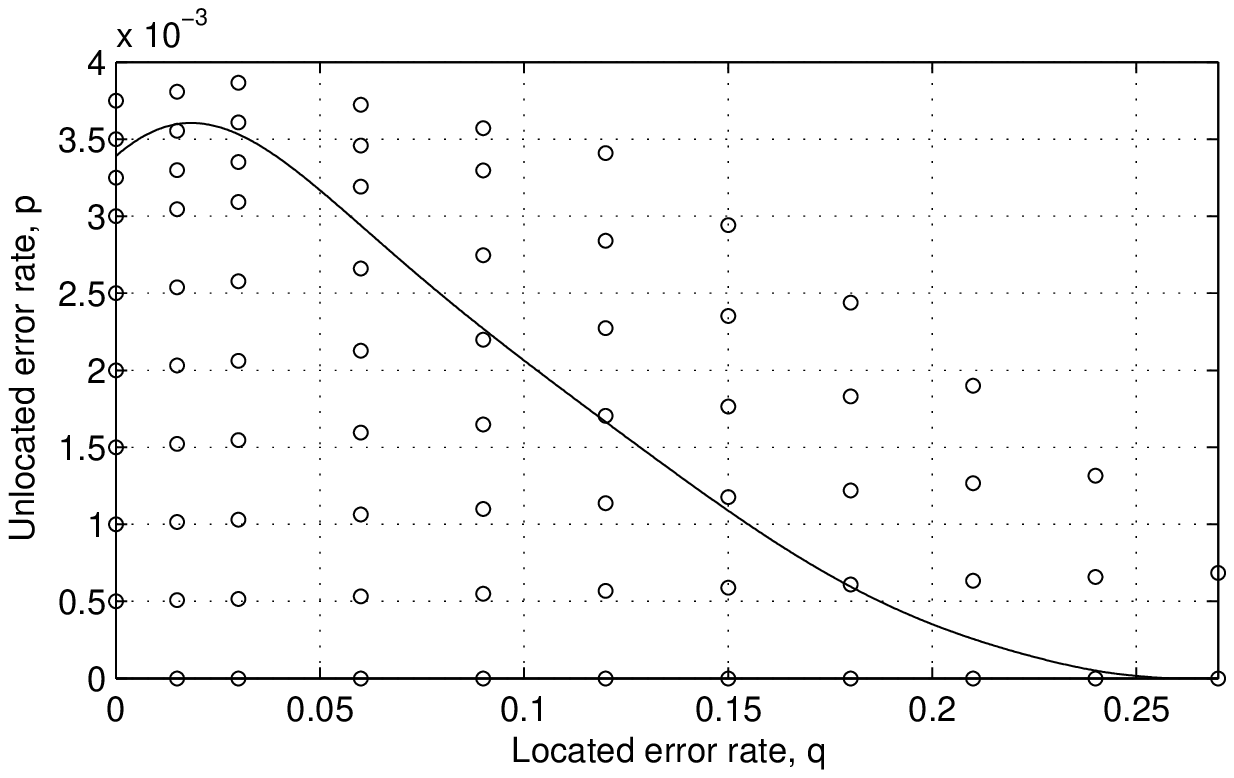}
\end{center}
\caption{Threshold region (below the solid line) for the deterministic protocol
  using the 23-qubit Golay code. Memory noise is disabled top, and
  enabled bottom. Dashed lines in upper figure show error due to
  finite sample size.}\label{fig:det23}
\end{figure}
Threshold regions for the simulations using the 23 qubit code are
shown in Figure~\ref{fig:det23}. The number of trials run per point
$(p_i,q_i)$ ranged from approximately $2\times 10^5$ to $4\times
10^7$. For the upper plot in Figure~\ref{fig:det23} we have estimated
the error in the threshold due to the finite sample size of the
simulations. This rough estimate of the error was obtained by
repeating the polynomial fitting a further 20 times, using the same
set of data $(P_i,Q_i)$, but subject to additional additive Gaussian
noise of standard deviation $(\sigma^P_i,\sigma^Q_i)$. The largest and
smallest values of the threshold obtained through this process are
plotted as the dashed lines. The estimated error for the other three
plots in Figures \ref{fig:det7} and \ref{fig:det23} is not shown, but
is smaller in these cases.

The threshold with respect to unlocated noise can be compared to
circuit-model thresholds obtained by other authors (keeping in mind
though that noise models and resource usage vary substantially between
different authors). Our best threshold for unlocated noise for the
four plots in Figures \ref{fig:det7} and \ref{fig:det23} is
approximately $8\times 10^{-3}$, for the $23$-qubit code with no
memory noise. This compares with a threshold of $3\times 10^{-3}$
obtained by Steane \cite{Steane03a}, $9 \times 10^{-3}$ by Reichardt
\cite{Reichardt04a}, and $3\times 10^{-2}$ by Knill \cite{Knill05a}.

A feature of our threshold plots worth noting is the dramatically
larger threshold for located noise (up to $0.25$ for the Golay code)
as compared to that of unlocated noise. Thus, the use of
post-selection in the protocol combined with a purpose-built decoding
routine has had a dramatic positive effect on the threshold for
unlocated noise.

Note also that all threshold regions in Figures \ref{fig:det7} and
\ref{fig:det23} show an unexpected feature: the threshold for
unlocated noise actually {\em improves} when a small amount of located
noise is added. Presumably, the presence of located noise converts
some crashes from unlocated to located, which are then more
efficiently dealt with by higher levels of concatenation. So, although
it would seem a somewhat absurd notion that adding noise should ever
improve the reliability of an error-correction protocol, such
behaviour in this case highlights how advantageous it can be to pass
information (i.e., crash locations) from one level to another in a
concatenated protocol.  Such behaviour appears somewhat similar to the
well-known phenomenon of stochastic resonance, whereby adding noise to
a system may in some circumstances actually improve the
signal-to-noise ratio in observations made on that system.

\subsection{Final Results}
\label{subsec:results}

In this subsection we give the final threshold results for optical
cluster-state quantum computing, with respect to the physical error
rates of our noise model.

Under $k$ layers of concatenation, our error-correction protocol
consists of one level of the optical cluster protocol concatenated
with $k-1$ levels of the deterministic protocol. Define the maps $f:
(\epsilon,\gamma)\rightarrow
(E(\epsilon,\gamma),\Gamma(\epsilon,\gamma))$ and $g: (p,q)
\rightarrow (P(p,q),Q(p,q))$, where $E$ and $\Gamma$ are the
polynomials obtained for the optical cluster protocol in Subsection
\ref{subsubsec:ocluster_results} and $P$ and $Q$ are the polynomials
obtained for the deterministic protocol in Subsection
\ref{subsubsec:det_results}. If $\epsilon$ is the depolarization
parameter and $\gamma$ is the photon loss rate (defined in
Section~\ref{sec:physical_setting}) then the unlocated and located
crash rates at level $k$ may be estimated by computing
$(g^{(k-1)}\circ f) (\epsilon,\gamma)$. If this tends to $(0,0)$ as
$k\rightarrow \infty$ then the physical noise rates
$(\epsilon,\gamma)$ are below the threshold.

Note that in deriving the results in this section, we are imagining
that the same code (either 7-qubit or 23-qubit) is used at every level
of concatenation. This need not be the case, and in general it is
possible to imagine a situation where the code choice is made
independently at each level.

\begin{figure}
\begin{center}
\epsfxsize=8.5cm \epsfbox{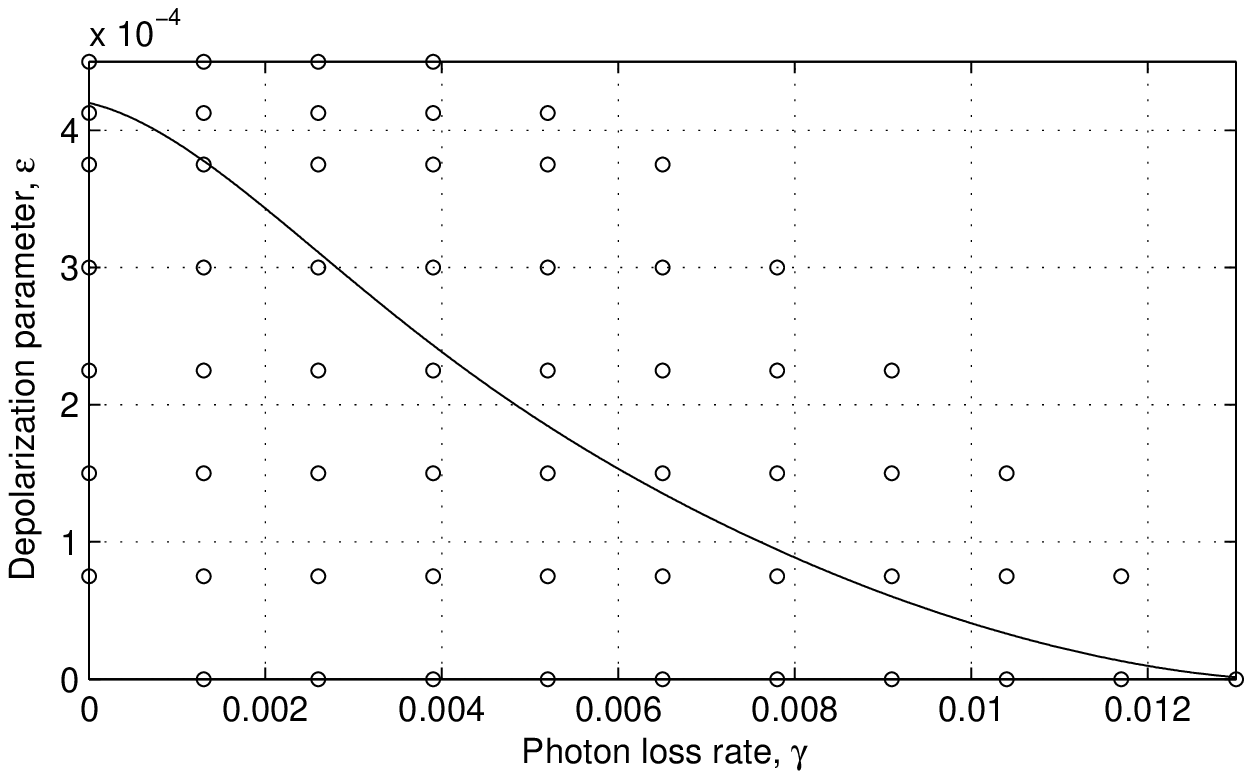}
\end{center}
\begin{center}
\epsfxsize=8.5cm \epsfbox{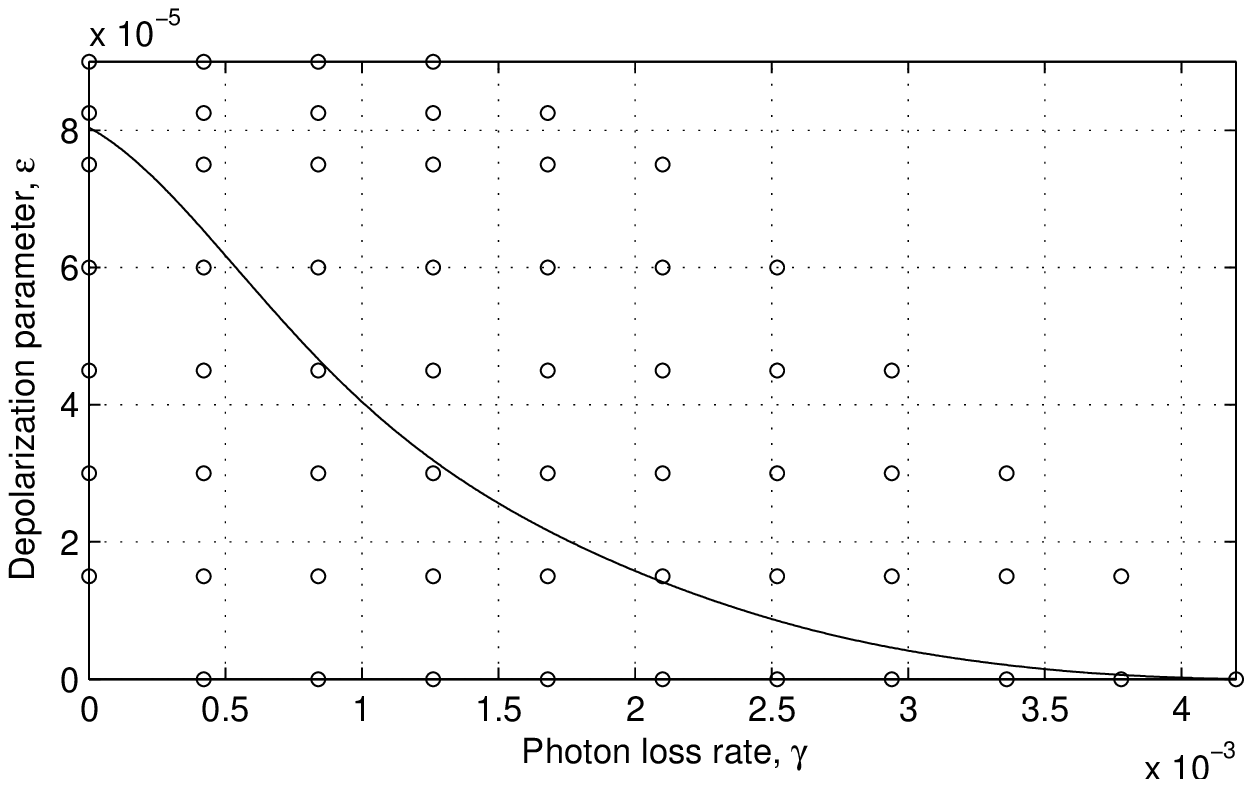}
\end{center}
\caption{Threshold region (below the solid line) for the optical cluster
  protocol using the 7-qubit Steane code. Memory noise is disabled
  top, and enabled bottom. Circles are located at the noise parameter
  values for which the cluster simulation was run.} \label{fig:oclus7}
\end{figure}
\begin{figure}
\begin{center}
\epsfxsize=8.5cm \epsfbox{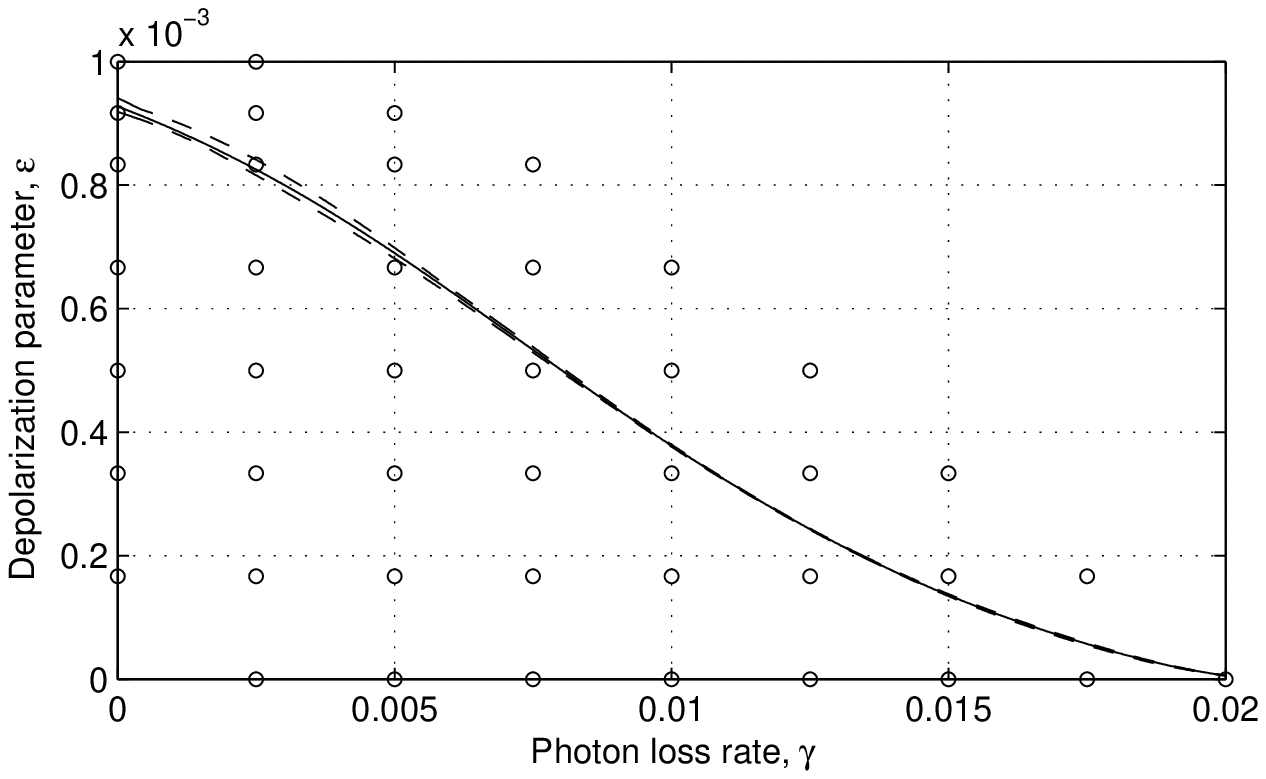}
\end{center}
\begin{center}
\epsfxsize=8.5cm \epsfbox{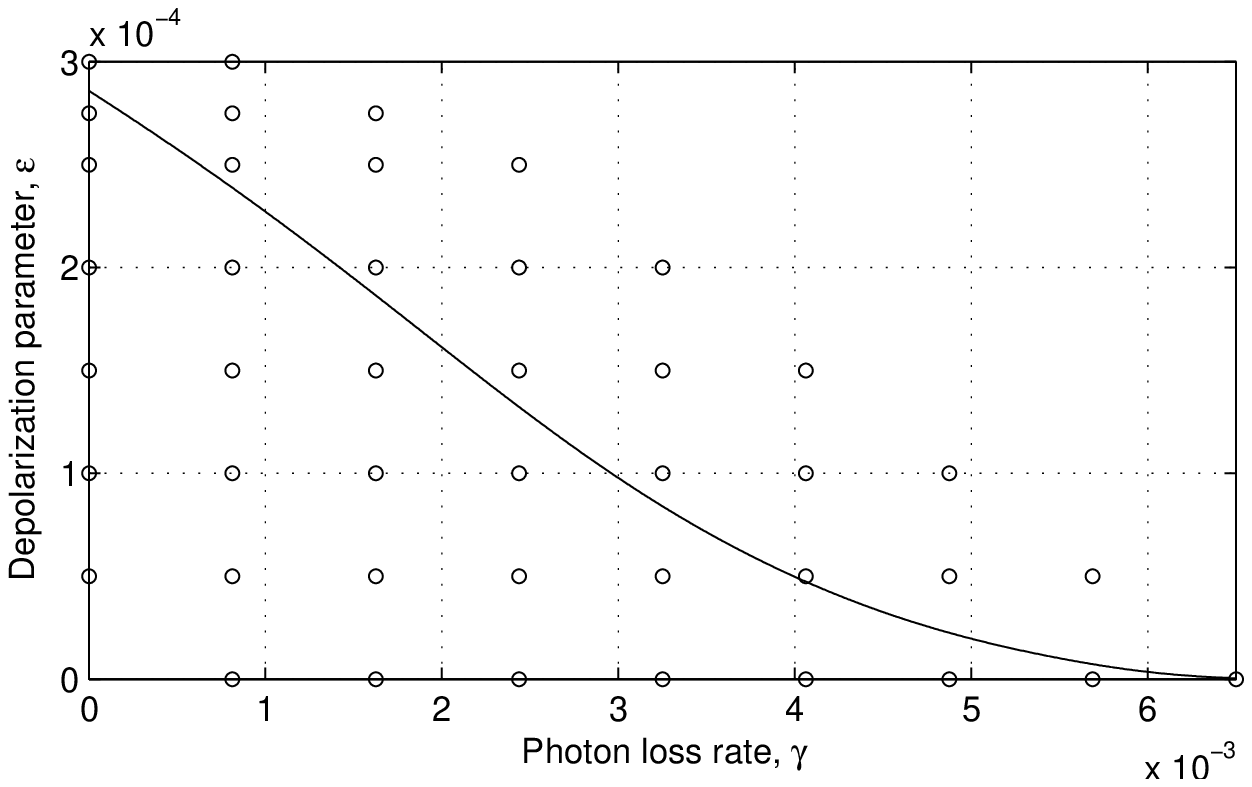}
\end{center}
\caption{Threshold region (below the solid line) for the optical cluster
  protocol using the 23-qubit Golay code. Memory noise is disabled
  top, and enabled bottom. Dashed lines in the upper figure show error
  due to finite sample size. }\label{fig:oclus23}
\end{figure}

The threshold regions using the $7$-qubit and $23$-qubit codes are
shown in Figures \ref{fig:oclus7} and \ref{fig:oclus23} respectively.
In the upper plot in Figure \ref{fig:oclus23} we have estimated the
error in the threshold due to the finite sample size of the
simulations, using a method similar to that of Subsection
\ref{subsubsec:det_results}. The estimated error is not shown in the
other three plots, but is smaller in these cases.

The best of the four thresholds is given by the $23$-qubit code with
no memory noise. In this case, the protocol can simultaneously protect
against a depolarization strength of $4\times 10^{-4}$ and photon loss
rate of $10^{-2}$, approximately. As expected, these values are poorer
than for the concatenated circuit-based protocol, due to the overhead
associated with clusterization of the optical protocol. We consider
these values encouraging, especially given the nondeterministic nature
of the optical two-qubit interactions.

\subsection{Resource usage}

In this subsection we perform a simple analysis of resource usage.
This analysis is performed for the Steane $7$-qubit code, and for a
particular physical noise rate. Ideally, a fuller analysis would
consider the (rather complex) question of how resource usage varies
with physical noise rates, code choice, and other variable aspects of
the protocol such as number of parallel fusions. However, the present
analysis is merely aimed at giving a very rough idea of resource
requirements.

For a measure of resource usage, we count the average number of Bell
pairs consumed per encoded operation. This measure can also be
considered as a rough indication of the usage of the other basic
operations (fusion gate, measurement, memory), since in the protocol
these operations are always very closely associated with Bell pair
creation and vice versa.

By ``per encoded operation'' in the description of the resource usage
measure above, we are referring to an operation at the highest level
of concatenation (that is, an actual logical gate of the computation
being carried out). Henceforth, we refer to such operations as
``computational operations''. The resource usage figure will thus
depend on the number of levels of concatenation. In turn, the number
of levels of concatenation required will depend on the desired level
of reliability of the final output of the computation, and the total
number of computational operations performed. For the sake of the
present analysis, let us define a ``reliable'' computation to be as
follows: with probability at least $\frac{1}{2}$ all computational
operations are crash-free (with respect to the highest level of
encoding). Assuming noise rates are below the threshold, adding more
levels of concatenation will give a lower probability of crash per
computational operation, and thus increase the maximum number of
computational operations allowed such that the output will be
reliable. If the total crash probability per computational operation
is $p_c$, then the output will be reliable if the number of
computational operations is less than
\begin{equation}
\frac{\log(\frac{1}{2})}{\log{(1-p_c)}}. \label{eq:maxlength}
\end{equation}

In Table \ref{table:resource}, the results of the analysis are shown,
for the $7$-qubit code with memory noise enabled. The chosen physical
noise parameters are
$(\epsilon,\gamma)$=$(4\times10^{-5},4\times10^{-4})$, corresponding
to a point roughly in the centre of the threshold region. Each row of
the table corresponds to a different number of levels of
concatenation. The effective rates of unknown and known crashes at
each level are shown in the columns $p$ and $q$. These values were
obtained by iterating the polynomials generated from the numerical
simulations\footnote{For the purposes of this analysis, we disallowed
further low-order terms in the polynomial that by the principles of
fault tolerance should be zero. This was done with the aim of
increasing the accuracy for very small parameter values.}. The maximum
computation length for each level was calculated from Equation
(\ref{eq:maxlength}) with $p_c=p+q$. The value for Bell pairs consumed
per computational operation, at a particular level $L$, is given by a
the number of Bell pairs consumed per error-correction step at level
$1$, multiplied by appropriate scale-up factors for each of the levels
$2,\dots,L$. The scale-up factor at some level $l$ is the expected
number of level $l-1$ error-correction steps used to implement a level
$l$ error-correction. These factors were estimated by the simulator in
a straightforward way (the details of the estimation procedure are not
given).

\begin{table}
\begin{tabular}{c|c|c|c|c}
Level & $p$ & $q$ & \parbox[c]{2.2cm}{Max.~reliable\\comp.~length} & \parbox[c]{2.5cm}{Bell pairs used per comp.~op.}\\
\hline%
1 & $0.00046$ & $0.0097$ & $68$ & $1.3\times 10^{8}$ \\
2 & $0.00022$ & $0.0027$ & $2.4\times 10^{2}$ & $1.5\times 10^{11}$ \\
3 & $4.4\times 10^{-5}$ & $0.00036$ & $1.7\times 10^{3}$ &  $9.3\times 10^{13}$ \\
4 & $1.5\times 10^{-6}$ & $9.9\times 10^{-6}$ & $6.1\times 10^4$ & $5\times 10^{16}$ \\
5 & $1.6\times 10^{-9}$ & $9.4\times 10^{-9}$ & $6.3\times 10^7$ & $2.6\times 10^{19}$ \\
6 & $1.9\times 10^{-15}$& $9.8\times 10^{-15}$&$5.9\times 10^{13}$ & $1.4\times 10^{22}$
  \end{tabular}
\caption{Estimated resource usage (number of Bell pairs consumed per
computational operation) as a function of concatenation level, for
noise parameters
$(\epsilon,\gamma)$=$(4\times10^{-5},4\times10^{-4})$, and using the
$7$-qubit code.
}\label{table:resource}
\end{table}

Thus, we see that to get a reliable computation consisting of a
significant amount of operations (say $10^9$), the protocol as it
stands has the very demanding requirement of approximately $10^{20}$
Bell pairs per operation. That this figure is so large can be partly
explained by our liberal use of post-selection in the various parts of
the protocol. Since our main aim in this paper is to find the
threshold for optical quantum computing, our protocol was designed
with optimization of the threshold the primary goal, and thus
optimization of resource usage was a lesser priority. A number of
simple modifications to the protocol would reduce the resource usage
by a few orders of magnitude at least, while only having a small
detrimental effect on the threshold. Such modifications would include
increasing the number of attempts per parallel fusion so that clusters
are discarded less often, and spreading cluster-building procedures
over more time steps so that smaller clusters are discarded if a step
fails. Nonetheless, resource usage is certainly a significant problem
both for our protocol and others (especially those that heavily rely
on post-selection such as \cite{Reichardt04a} and \cite{Knill05a}).

\section{Conclusion}
\label{sec:conclusion}

In this paper we've done a detailed numerical investigation of the
fault-tolerant threshold for optical cluster-state quantum computing.
Our work considers a noise model which allows for both photon loss and
depolarizing noise.  Depolarizing noise is used as a general proxy for
all types of local noise other than photon loss, and standard results
in the theory of error-correction ensure that the ability to protect
against depolarization ensure the ability to protect against other
types of noise, including dephasing, amplitude damping, etc.

Our main result has been a \emph{threshold region} of allowed pairs of
values for the photon loss and depolarizing noise.  Roughly speaking,
our results show that scalable optical quantum computing is possible
in the combined presence of both noise types, provided that the loss
probability is $< 3 \times 10^{-3}$ and the depolarization probability
is $< 10^{-4}$.  To achieve such threshold values requires very
substantial overheads in order to accurately perform long
computations.  Future work will need to not only improve the
threshold, but also reduce the overhead required to do fault-tolerant
computation, improve the accuracy of the noise model used in
simulations, and address the pseudothreshold phenomenon identified
in~\cite{Svore05a,Svore05b}.

Our noise model is in contrast to previous investigations of the
threshold for optical quantum computing, which have focused on the
case in which photon loss is the \emph{sole} source of noise.  While
photon loss will certainly be an important source of noise in real
implementations, other sources of noise such as dephasing will also be
present (at lower levels), and techniques which protect solely against
photon loss will have the effect of greatly amplifying those other
sources of noise.  Thus, while the earlier loss-only thresholds are of
considerable theoretical interest, they do not provide physically
meaningful thresholds.

We note that our threshold results might be applicable to
implementations of quantum computing other than linear optics -- in
particular to any scheme that contains nondeterministic two-qubit
interactions, loss noise and depolarization noise. For example, in the
scheme by Barret and Kok \cite{Barrett04a} for quantum computing with
matter qubits, two-qubit interactions are nondeterministic, with a
heralded failure rate of 50\%. In analogy to photon loss, the scheme
can also exhibit ``loss'' when an atom jumps out of the qubit space
into a higher energy level. It is likely that our threshold results
would agree at least qualitatively with the thresholds for such a
system.

\acknowledgments

Many thanks to Maries Hankel of the CCMS computing facility. Thanks to
Ike Chuang, Jennifer Dodd, Steve Flammia, and Tim Ralph for helpful
discussions. Thanks to Peiter Kok for helpful correspondence. We thank
Bryan Eastin and Steve Flammia for their circuit drawing package
\emph{Qcircuit}.

\appendix
\section{Telecorrection}
\label{sec:telecorrection}

This appendix presents the idea of the telecorrector in
a simple form, without the baggage of clusters and nondeterminism.
Although use of the telecorrector arises naturally from the cluster
state model, there are good reasons to consider it in the circuit
model as well. First, it provides a different way of thinking about
quantum error correction: most of the difficulty of a fault tolerant
round of quantum error correction can be reduced to the creation of a
single $2n$-qubit resource state ($n$ being the size of the code).
This is in contrast to the normal requirements of an error correction
round -- the creation of a variable number (at least four) copies of
an $n$-qubit ancilla state. While we shall consider a particular way
of generating the telecorrector, based on a teleported Steane syndrome
extraction circuit, it is an interesting open problem to consider
better methods for creation.

The second reason for considering telecorrection in the circuit model
is for practical use in our simulations. Our deterministic error
correction protocol, used for the second and higher levels of
concatenation, uses circuit-model telecorrection instead of the
standard Steane approach. The benefit is an improved threshold, due to
the ability to post-select for agreeing syndromes and against located
noise types during telecorrector creation.

Note that the idea of combining error correction and teleportation has
been used previously by Knill \cite{Knill05a}, however the details of
our telecorrection procedure and Knill's procedure differ
significantly in the details.

We now derive a circuit for fault-tolerant telecorrection. Begin with
the following circuit for Steane's repeated syndrome extraction:
\begin{equation} \label{eq:steane2}
\epsfxsize=6.5cm \epsfbox{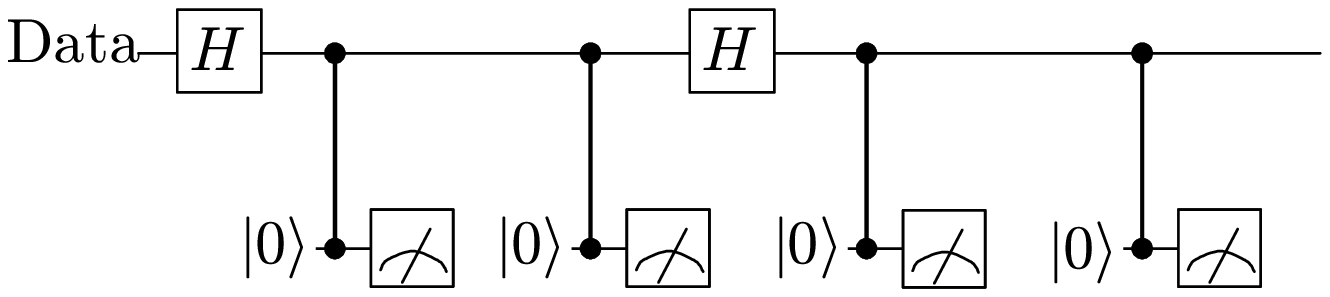}.
\end{equation}
In the circuit, wires and gates represent encoded qubits and encoded
operations respectively, and $|0\rangle$ represents Steane's
fault-tolerant ancilla creation circuit. The circuit performs two $Z$
syndrome extractions followed by two $X$ syndromes, and generalizes to
more than two extractions of each in the obvious way.

We replace each of the encoded Hadamard operations in
Circuit~(\ref{eq:steane2}) by the transport circuit of Equation
(\ref{eq:transport}), to give
\begin{equation} \label{eq:telesteane}
\epsfxsize=6.5cm \epsfbox{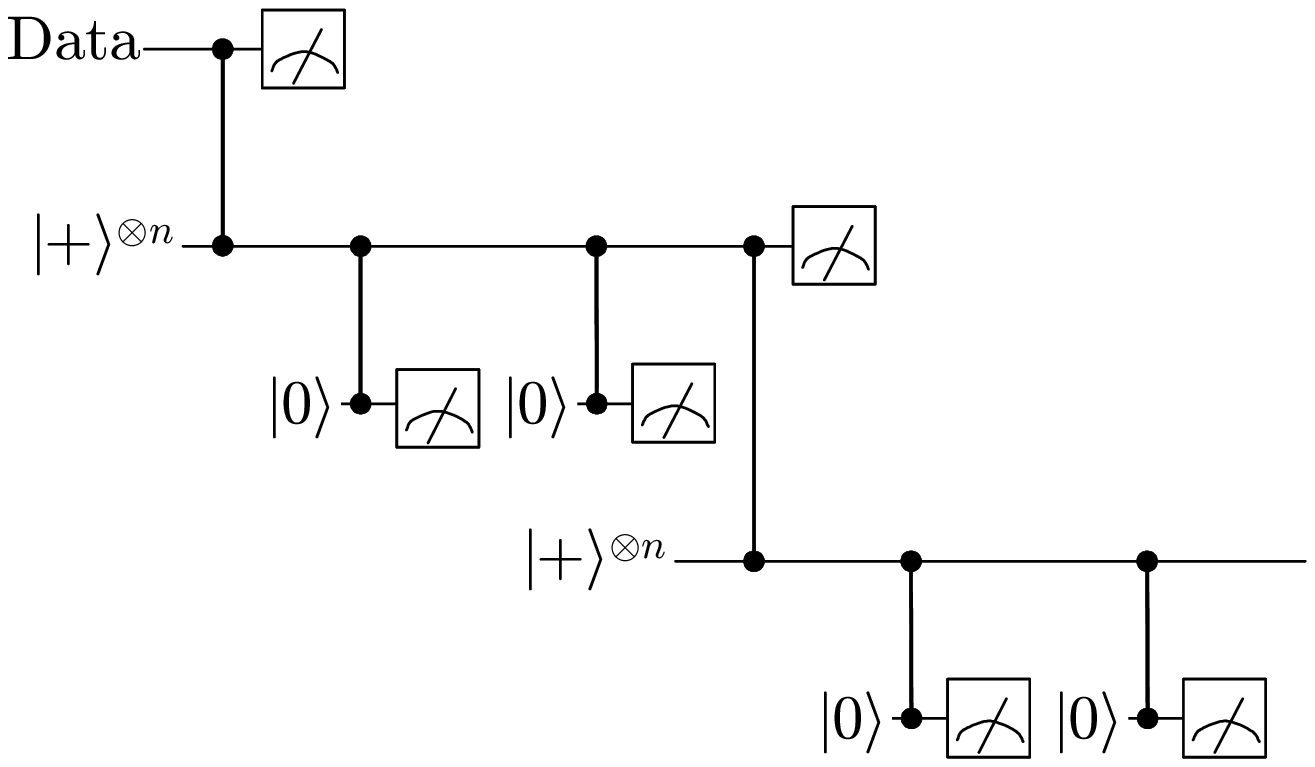}
\end{equation}
where we have omitted showing the necessary classical feed-forward
associated with the transport steps. We commute various operations to
finally give
\begin{equation} \label{eq:telecorrector1}
\epsfxsize=5cm \epsfbox{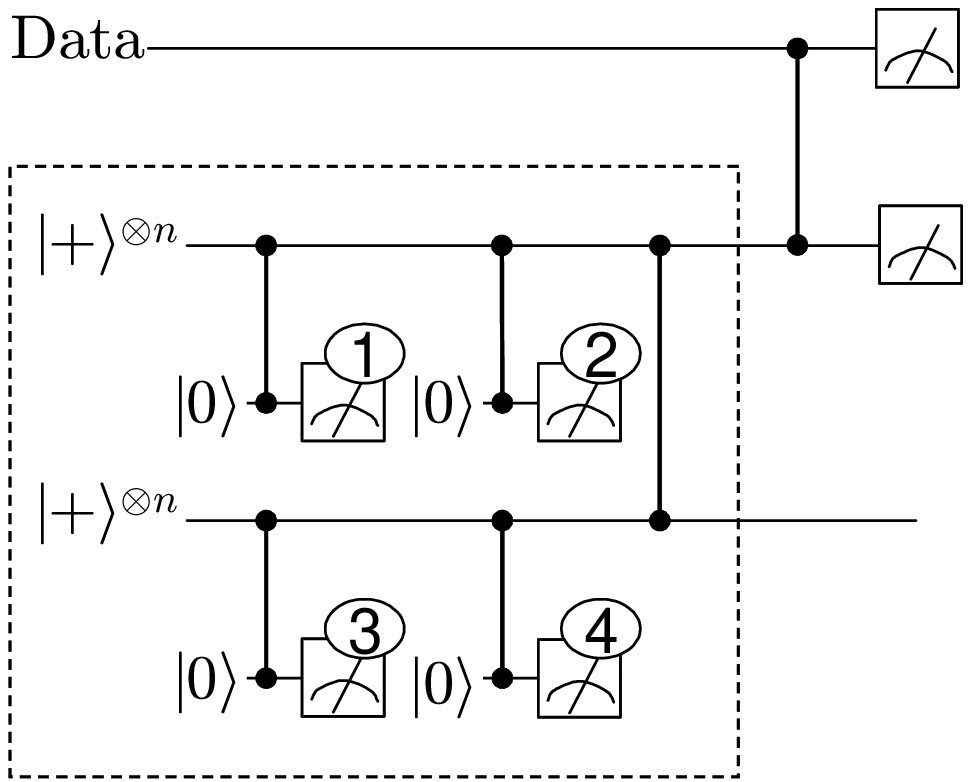}.
\end{equation}
The dashed box encloses the telecorrector creation circuit.
Measurements 1 and 2 correspond to $Z$ syndrome measurements of the
original circuit, and measurements 3 and 4 correspond to $X$ syndrome
measurements. However, these measurements do not directly give the
syndromes of the input data, since they must be adjusted due to the
output of the transport-circuit measurements. The details of this will
be derived below. Note however that we can determine if the syndromes
of like type will agree, and post-select for this, before the
telecorrector has interacted with the data.

To understand exactly what state the telecorrector is, we now consider
the evolution of Circuit (\ref{eq:telecorrector1}), in the case of
noise-free operations. The telecorrector creation circuit begins with
the state $|+\rangle^{\otimes n} \otimes |+\rangle^{\otimes n}$. We
note that the state $|+\rangle^{\otimes n}$ can be written, without
normalization, as
\begin{equation} \label{eq:plustensor}
|+\rangle^{\otimes n} = \sum_{s=0}^{2^{(n-1)/2}-1}
\vec{X}(s)|+\rangle_L,
\end{equation}
where $s$ labels $X$-syndromes of the code (we are assuming the code
encodes one qubit, hence there are $2^{(n-1)/2}$ $X$-syndromes),
$\vec{X}(s)$ is some tensor product of $X$s and $I$s having syndrome
$s$, and $|+\rangle_L$ is the encoded $|+\rangle$ state. Each
$|+\rangle^{\otimes n}$ in the circuit undergoes two $X$ syndrome
extractions, each consisting of a controlled phase with an encoded
$|0\rangle$ ancilla and subsequent measurement. The first $X$ syndrome
extraction performed on a $|+\rangle^{\otimes n}$ randomly collapses
it to one of the terms in Equation~(\ref{eq:plustensor}). In the
noise-free case, the second syndrome extraction has no effect. Thus,
the state of the telecorrector after all syndrome extractions, but
before the controlled phase connecting the two halves, is
\begin{equation}
\left(\vec{X}(s_z)|+\rangle_L\right)\otimes\left(\vec{X}(s_x)|+\rangle_L\right),
\end{equation}
where $s_z$ and $s_x$ represent the syndrome measurement results from
the top and bottom halves of the circuit respectively.

Next we apply the controlled phase between the two halves of the
telecorrector creation circuit. This gives the state
\begin{equation}
\left(\vec{X}(s_z)\vec{Z}(s_x)\otimes \vec{X}(s_x)\vec{Z}(s_z)\right)
\Big| \Qcircuit[1em] @R=1em @C=1em { \node{} & \node{} \link{0}{-1} }
\Big\rangle,
\end{equation}
where we have commuted the controlled phase through the $X$
operations, and the ket is the encoded two-node cluster state. Thus, a
noise-free telecorrector state is an encoded two-node cluster state,
up to known Pauli operations.

We now consider the remaining operations in
Circuit~(\ref{eq:telecorrector1}) that complete the telecorrection of
the data. The controlled phase between the telecorrector and data
gives the state
\begin{equation}
\left(\vec{Z}(s_z) \otimes \vec{X}(s_z)\vec{Z}(s_x)\otimes
\vec{X}(s_x)\vec{Z}(s_z)\right) \Big| \Qcircuit[1em] @R=1em @C=1em
{\psi & \node{}\link{0}{-1} & \node{} \link{0}{-1} } \Big\rangle,
\end{equation}
where the ket is the state (on encoded qubits) obtained by applying a
controlled phase between the data state, denoted $\psi$, and a
two-node cluster state.

The final two measurements are then performed. These are encoded
$X$-basis measurements on the data and one half of the telecorrector.
An encoded $X$-basis measurement is performed by measuring each
physical qubit in the $X$ basis, adjusting the measurement results to
remove the effects of known Pauli operations ($\vec{Z}(s_z)$ in the
case of the measurement of the data, and $\vec{Z}(s_x)$ in the case of
the measurement of the top half of the telecorrector), and performing
classical error correction on the resulting bit string. For the codes
we consider, the measurement outcome is $0$ if the corrected bit
string has even weight, and $1$ otherwise. The corrections performed
during the two encoded measurements have the effect of eliminating any
errors present in the input state, and also certain errors introduced
by telecorrector creation, subject to the weight of those errors being
not too large.

Let the measurement result on the data and telecorrector be $m_1$ and
$m_2$ respectively. Then, the output state is
\begin{equation}
\vec{X}(s_x)\vec{Z}(s_z)(Z^{\otimes n})^{m_1}(X^{\otimes n})^{m_2}
|\psi\rangle,
\end{equation}
to which we apply the appropriate Pauli operators, giving the final
output of the telecorrector, the error-corrected version of the state
$|\psi\rangle$.

Finally, note that the following straightforward modification to the
telecorrection circuit,
\begin{equation} \label{eq:telecorrector2}
\epsfxsize=5cm \epsfbox{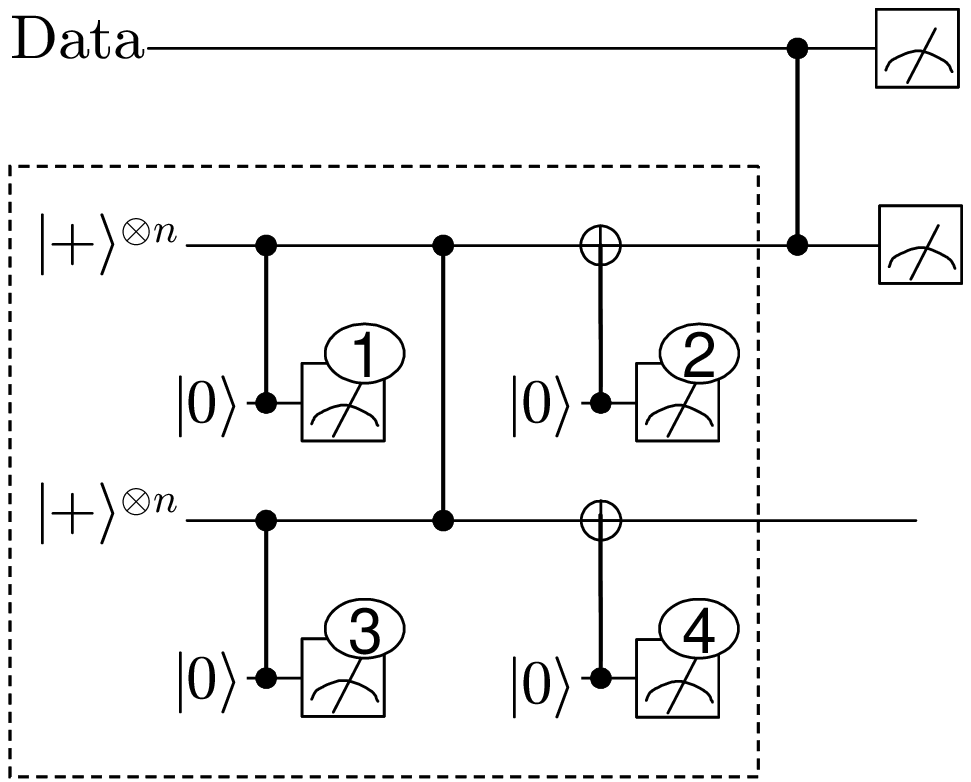},
\end{equation}
provides an improved noise-threshold performance compared with
Circuit~(\ref{eq:telecorrector1}). In
Circuit~(\ref{eq:telecorrector2}), measurements 1 and 4 correspond to
$Z$ syndrome extraction in Steane's protocol, and measurements 2 and 3
correspond to $X$ syndrome extraction. The circuit has the property
that the post-selection for preagreeing syndromes will eliminate a
larger class of errors than is the case for
Circuit~(\ref{eq:telecorrector1}). For example, $X$ errors which
propagate from either ancilla 1 and 3 to become $Z$ errors on the
telecorrector will very likely cause syndromes $2$ or $4$ to disagree
with $3$ or $1$ respectively. Also, certain types of noise caused by a
failed controlled phase between the two halves of the telecorrector
will also cause disagreeing syndromes.


\end{document}